\documentclass[final,spanish,color]{article}
\usepackage[spanish, es-tabla]{babel}
\usepackage{geometry} 
\usepackage[utf8]{inputenc}
\usepackage[centertags]{amsmath}
\usepackage{amssymb}
\usepackage{amsthm}
\usepackage{amscd}
\usepackage{makeidx}
\usepackage{delarray}
\usepackage{amsmath}
\usepackage{colortbl}
\usepackage{anysize}
\usepackage{float}
\usepackage{graphicx}
\usepackage{enumerate}
\usepackage{latexsym,amstext,amscd}
\usepackage{fancyhdr}
\usepackage{amsfonts}
\usepackage{caption}
\usepackage{gensymb}
\usepackage{anysize}
\usepackage{capt-of}
\usepackage{multicol}
\usepackage{multirow}
\usepackage{sidecap}
\usepackage[font=small,skip=0pt]{caption}
\usepackage{subcaption}
\usepackage{cite}
\usepackage{pdfpages}
\usepackage{morefloats}
\usepackage[toc,page]{appendix}
\usepackage[labelfont=bf]{caption}

\usepackage{array}
\usepackage{emptypage}

\makeatletter
\def\@makeappendixhead#1{%
	\null\vfill%
	{\parindent \z@ \centering \normalfont
		\ifnum \c@secnumdepth >\m@ne
		\if@mainmatter
		\huge\bfseries \@chapapp\space \thechapter
		\par\nobreak
		\vskip 20\p@
		\fi
		\fi
		\interlinepenalty\@M
		\Huge \bfseries #1\par\nobreak
		\vfill
		\clearpage
}}
\g@addto@macro\appendices{\let\@makechapterhead\@makeappendixhead}
\makeatother

\addto\captionsspanish{%

}

\begin{document}

\begin{center}
 \textbf {\Large{Brain MRI study for glioma segmentation using convolutional neural networks and original post-processing techniques with low computational demand}}
\vspace{0.5cm}

José Gerardo Suárez-García$^{1a}$\footnote{The author JGSG was supported by the National Council of Science and Technology (CONACYT) to carry out this work, through a scholarship for postgraduate studies.}, Javier Miguel Hernández-López$^{1}$, Eduardo Moreno-Barbosa$^{1}$ and Benito de Celis-Alonso$^{1}$

$^1$Faculty of Physics and Mathematics, Benemérita Universidad Autónoma de Puebla (BUAP), Puebla, Puebla, México

$^a$gsuarez.biofis@gmail.com

\end{center}

\begin{abstract}
Gliomas are brain tumors composed of different highly heterogeneous histological subregions. Image analysis techniques to identify relevant tumor substructures have high potential for improving patient diagnosis, treatment and prognosis. However, due to the high heterogeneity of gliomas, the segmentation task is currently a major challenge in the field of medical image analysis. In the present work, the database of the Brain Tumor Segmentation (BraTS) Challenge 2018, composed of multimodal MRI scans of gliomas, was studied. A segmentation methodology based on the design and application of convolutional neural networks (CNNs) combined with original post-processing techniques with low computational demand was proposed. The post-processing techniques were the main responsible for the results obtained in the segmentations. The segmented regions were the whole tumor, the tumor core, and the enhancing tumor core, obtaining averaged Dice coefficients equal to 0.8934, 0.8376, and 0.8113, respectively. These results reached the state of the art in glioma segmentation determined by the winners of the challenge.
\end{abstract}

\section{Introduction}

Gliomas are brain tumors with different degrees of aggressiveness and are composed of highly heterogeneous subregions in appearance, shape and histology \cite{prot_seg}.
Heterogeneity can be appreciated in its image phenotype (appearance and shape), since the subregions are described by different intensity profiles scattered across MRIs, reflecting the different biological properties of the tumor. 
The correct choice of the patient treatment strategy and the evaluation of the progression of the disease are essential for its prognosis. Consequently, there is currently a need to develop image analysis techniques to accurately and reproducibly identify relevant tumor substructures. This would improve the diagnosis, planning and follow-up of the patient.
Due to the heterogeneity of gliomas, their segmentation is a challenge important in the field of medical image analysis \cite{cc2}. The goal of this work was to create a automatic segmentation methodology of the following subregions of gliomas: the whole tumor, the tumor core and the enhancing tumor core.
Taking as a reference the works presented in the BraTS 2018 and 2019 challenge, the most used mathematical model for segmentation was the Convolutional Neural Networks (CNNs). These networks are created specifically for image analysis. Its design aims to mimic the mechanism of the visual cortex of mammals, which is assumed to be formed by groups of ordered and specialized neurons for object recognition, starting from the simplest features to the most complex patterns \cite{cnn_intro}. One of the advantages of CNNs is that they automatically learn the necessary image features, without the need to be entered by the user. CNNs have been applied to solve different problems such as the classification of brain tumors \cite{ej1_cnn}, detection of skin lesions \cite{ej2_cnn}, detection of diabetes through images of heart rhythm frequencies \cite{ ej3_cnn}, breast cancer detection \cite{ej4_cnn}, COVID-19 detection through X-ray images \cite{ej5_cnn}, among many others.
The use of cascade methodologies stands out among the works of the challenge. These begin by differentiating healthy tissue from tumor tissue, and subsequently segment the different subregions of the tumor \cite{ej2_cnn}. Semantic segmentations have also been done, in which a set of voxels are labeled at the same time after applying a neural network to them \cite{semantic}. Unlike semantic segmentation, voxel-by-voxel segmentations have been performed, where one voxel per occasion is labeled \cite{pixel-wise}. In particular, a pre-trained CNN called U-Net has been frequently applied. This is a network architecture for fast and accurate image segmentation. It consists of a contraction structure (encoder) to acquire the context, an expansion structure (decoder) for precise localization, and symmetric concatenations for auto-context generation. The U-Net outperformed the prior best method (a sliding-window convolutional network) on the ISBI challenge for segmentation of neuronal structures in electron microscopic stacks, through training with a relatively small number of images \cite{unet}. This pre-trained network has been used through transfer learning and its architecture has served as a reference to create new networks to train. Other developed CNNs are called DeepMedic \cite{crf}, V-net \cite{vnet}, SegNet \cite{segnet}, ResNet \cite{resnet} and DenseNet \cite{densenet}.
In other works, the success of the segmentations depended almost exclusively on the capacity of the trained CNNs. At present, this was mainly determined by different original post-processing techniques proposed. This was motivated by hardware limitations that did not allow the development of complex CNNs as in other works. Then, based on the results of the CNNs, different techniques were incorporated to improve the segmentations and thus reach the state of the art.

\section{Methodology}

	In the present work, the database of the Brain Tumor Segmentation (BraTS) Challenge 2018, composed of multimodal MRI scans of gliomas, was studied \cite{brats2, brats5, prot_seg,brats_falt}. Different neural networks were created and combined to label individual voxels from three MRI modalities: T\textsubscript{1Gd}, T\textsubscript{2} and FLAIR, considering the following four possible labels: healthy tissue (TS), peritumoral edema (ED), necrotic and non-enhancing tumor core (NCR/TI), and enhancing tumor core (NA). Subsequently and during three stages, three subregions were segmented: whole tumor (ED, NCR/TI and NA labels) at stage 1, tumor core (NCR/TI and NA labels) at stage 2, and enhancing tumor core (NA label) in stage 3.

\subsection{Data division}

The neural networks studied a set of two-dimensional patches of size \emph{d}$\times$\emph{d} (where $d$ is an odd number) extracted from the MRI, whose voxel located in its center would be labeled. The database for \emph{The multimodal brain tumor segmentation challenge (2018)} was randomly divided into three subsets: training, validation and testing. For the training subset, 175 gliomas were chosen, for the validation subset 39 and for the test subset also 39. The neural networks studied a set of two-dimensional patches of size \emph{d}$\times$\emph{d} (where $d$ was an odd number) extracted from MRIs. The voxels located in the center of the patches were labeled.

\subsection{Stage 1}

The goal of stage 1 was to create a CNN for whole tumor segmentation by differentiating between voxels corresponding to healthy tissue (labeled TS) and voxels corresponding to the whole tumor (labeled ED, NA, or NCR/TI). The TC label was associated with the whole tumor. Thus, of the total number of patches used to train the network, one half had a TS label and the other half had a TC label (Table \ref{parches_etapa1}), chosen randomly from all the possible patches of interest extracted from the 175 training gliomas . Similarly, of the total number of patches used to validate the network, one half had the TS label and the other half TC, randomly chosen from all the possible patches of interest extracted from the 39 gliomas of validation.

\begin{table}[ht]
	\centering
	
	\begin{tabular}{ccc}
		\hline 
	\textbf{Label}  & \textbf{Training patches}  &	  \textbf{Validation patches}     \\				
		\hline  
		TS & 999,999    &	49,998		\\	
		\hline 
ED &		333,333     &	16,666	\\		
		\hline 
	NA &	333,333   &	16,666		\\	
		\hline 
	NCR/TI &	333,333    &	16,666	\\
		\hline 
		Total & 1,999,998 & 99,996 \\
		\hline
	\end{tabular}
	\caption[Training and validation patches for stage 1]{{\bf Training and validation patches for stage 1.} The total number of patches and their respective labels are indicated. The ED, NA, and NCRT/TI labels conformed to the TC label.}
	\label{parches_etapa1}
\end{table}

\begin{figure}[ht] 
	
	\begin{center} 
		\includegraphics[scale=0.45]{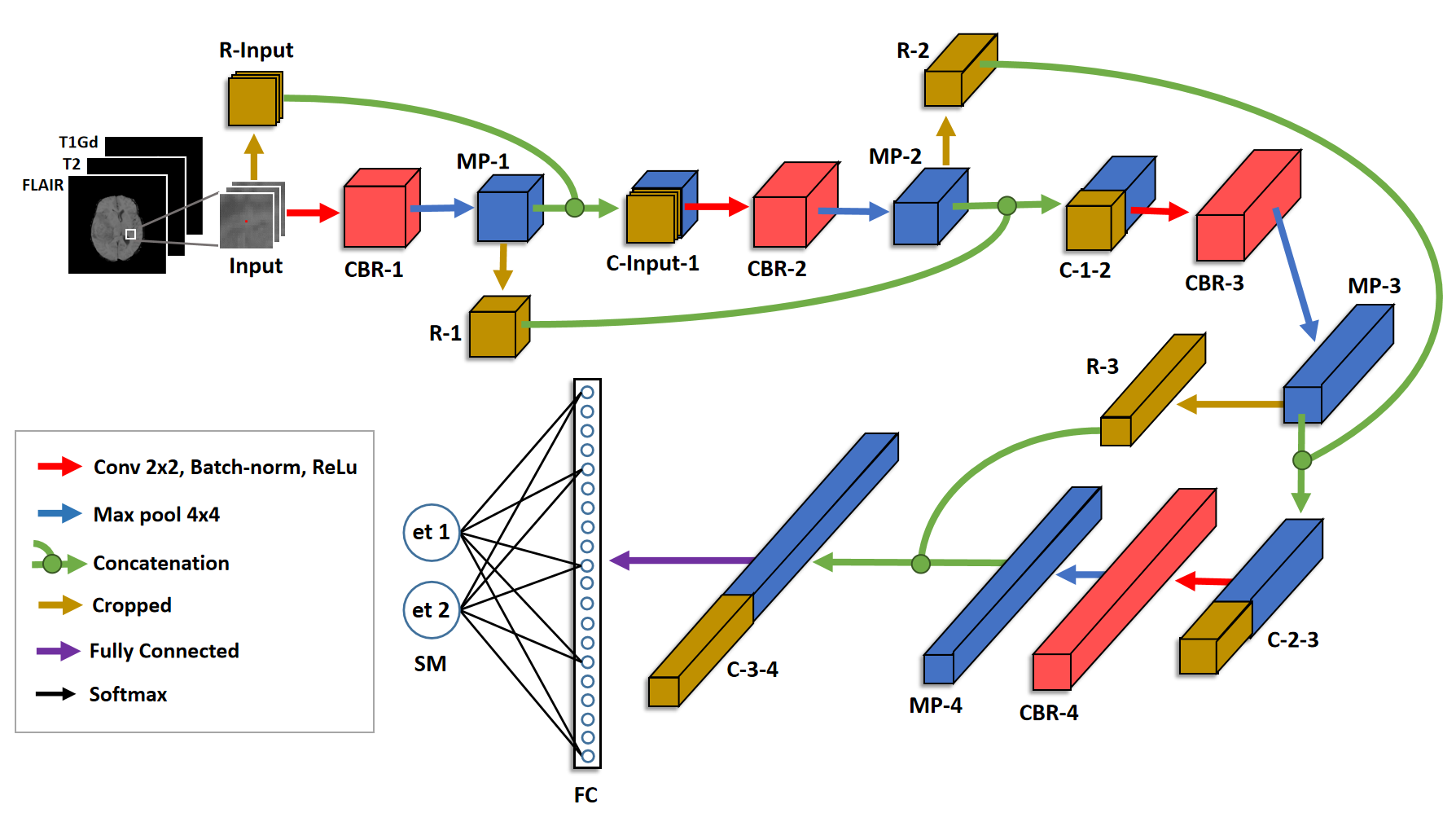} 
		
	\end{center} 
	\caption[Stage 1 CNN architecture]{{\bf Stage 1 CNN architecture.} The network input is shown as a patch of 15$\times$15$\times$3 voxels. Only the central voxel of the patch was labeled after applying the CNN. At the end of the CNN, the softmax function (SM) was used to obtain the outputs et 1 and et 2 corresponding to the TC and TS labels, respectively.}
	\label{red_1} 
\end{figure}

The CNN architecture is shown in Fig. \ref{red_1}. Considering, for example, the axial plane of an MRI volume, a network input (Input) was formed by three two-dimensional patches with a size equal to 15$\times$15 voxels. Each of these patches corresponded to the same brain region seen in each of the three MRI modalities T\textsubscript{1Gd}, T\textsubscript{2}, and FLAIR. The network inputs passed through different layers. The first of them indicated as CBR-1 was formed by a convolution layer, batch normalization and activation through the ReLu function. The number of convolution filters was equal to 32, with a size of 5$\times$5, with stride equal to 1 and padding such that their output was the same size as the input. Subsequently, a maxpooling layer (MP-1) with a size of 4$\times$4 and stride equal to 1 was applied. Then, a concatenation between the output of the maxpooling layer and a cropped of the network input patch (R-Input) was done. Since the maxpooling output had a size of 12$\times$12$\times$32, so in order to perform the concatenation, the input patch of size 15$\times$15$\times$3 was cropped to a size of 12$\times $12$\times$3 with the same center. So this cropped patch was concatenated with the maxpooling output along the third dimension to form a volume of size 12$\times$12$\times$35 (C-Input-1).

The above was repeated in a similar way three times with some variations. The number of convolution filters increased to 64, 128 and 256 respectively. Subsequent concatenations were made between the outputs of a maxpooling and a patch cropped from the output of the previous maxpooling, with the same center and using the appropriate dimensions to make the concatenation along the third dimension. After the last concatenation a volume of size 3$\times$3$\times$384 was obtained which was fully connected to a layer of 50 hidden nodes. In this layer, a dropout of 0.8 was used and the softmax function was applied with two outputs, each associated with the two possible labels TS and TC, respectively. The meaning of the outputs can be interpreted as the probability that the voxel was correctly associated with the respective label, each varying from 0 to 1 and both adding up to 1. Henceforth the outputs of the last layer were called scores. Details about the size of the outputs of each of the layers are shown in Table \ref{tabla3}. Since the sum of the two scores was equal to 1, only the score associated with the TC label was used. Therefore, when the score was greater than or equal to a threshold, then the voxel was labeled TC, otherwise it was labeled TS. The network was trained 100 epochs, using a minibatch equal to 32 and using the optimizer known as Adam, with a constant learning range equal to 10$^{-4}$. A total of 17 thresholds corresponding to scores from 0.15 ro 0.95 with steps of 0.05 were studied.  

\begin{table}[ht]
	\centering
	
	\begin{tabular}{cccc}
		\hline 
		\textbf{Layer}  & \textbf{Output size}   & \textbf{Layer}  & \textbf{Output size}    \\				
		\hline  
		Input & 15$\times$15$\times$3 & 	MP-3 &  6$\times$6$\times$128 	\\	
		\hline 
		CBR-1 & 15$\times$15$\times$32 &  R-2  & 6$\times$6$\times$64	\\	
		\hline 
		MP-1 & 12$\times$12$\times$32 & C-2-3   &	6$\times$6$\times$192 \\	
		\hline 
		R-Ipunt & 12$\times$12$\times$3 &  CBR-4  & 6$\times$6$\times$256	\\	 
		\hline 
		C-Input-1 & 12$\times$12$\times$35 & MP-4   &	3$\times$3$\times$256 \\	
		\hline 
		CBR-2 &12$\times$12$\times$64 & R-3   & 3$\times$3$\times$128	\\	
		\hline 
		MP-2 & 9$\times$9$\times$64 & C-3-4   & 3$\times$3$\times$384	\\	
		\hline 
		R-1 & 9$\times$9$\times$32 & FC   & 1$\times$1$\times$50	\\	
		\hline 
		C-1-2 & 9$\times$9$\times$96 &  SM & 1$\times$1$\times$2	\\	
		\hline 
		CBR-3 & 9$\times$9$\times$128 &   &	\\	
		\hline 
	
\end{tabular}
\caption[Layers and output sizes]{{\bf Layers and output sizes.} For the stage 1 CNN, information about the size of the output volumes of each of its layers is displayed.}
\label{tabla3}
\end{table}

\subsubsection{Construction and combination of different CNNs}

Three CNNs were built with the same architecture, studying the same voxels of interest, but using as inputs the patches obtained in the axial, coronal and sagittal planes, respectively. The score assigned to a voxel was equal to the average of the three scores obtained after applying each of the three mentioned CNNs.

\subsubsection{Score volumes}

\begin{figure}[ht] 
	
	\begin{center} 
		\includegraphics[scale=0.5]{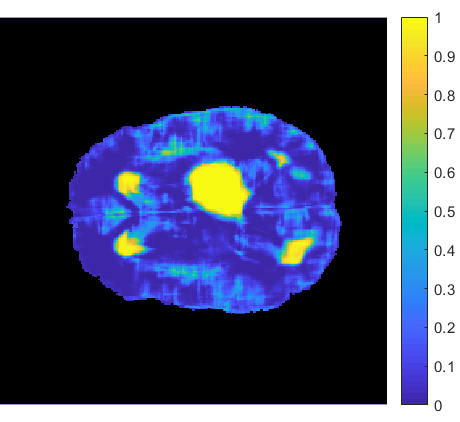} 
		
	\end{center} 
	\caption[Score volume for stage 1]{{\bf Score volume for stage 1.} Volume created after applying the three CNNs to all voxels and combining their three scores as the average of them. The scale indicates the probability that a voxel had the TC label.}
	\label{vol_punt} 
\end{figure}

After obtaining the scores of all the voxels that formed a complete volume, the so-called score volumes were created (Fig. \ref{vol_punt}). From these volumes it was observed that usually sets of voxels far away and isolated from the real segmentation and whose correct label was TS, were erroneously included in the segmentation made of the complete tumor. In order to correct these errors, original post-processing techniques called \emph{center} and \emph{connected neighbors} were developed.

\subsubsection{Center}

\begin{figure}[ht] 
	
	\begin{center} 
		\includegraphics[scale=0.15]{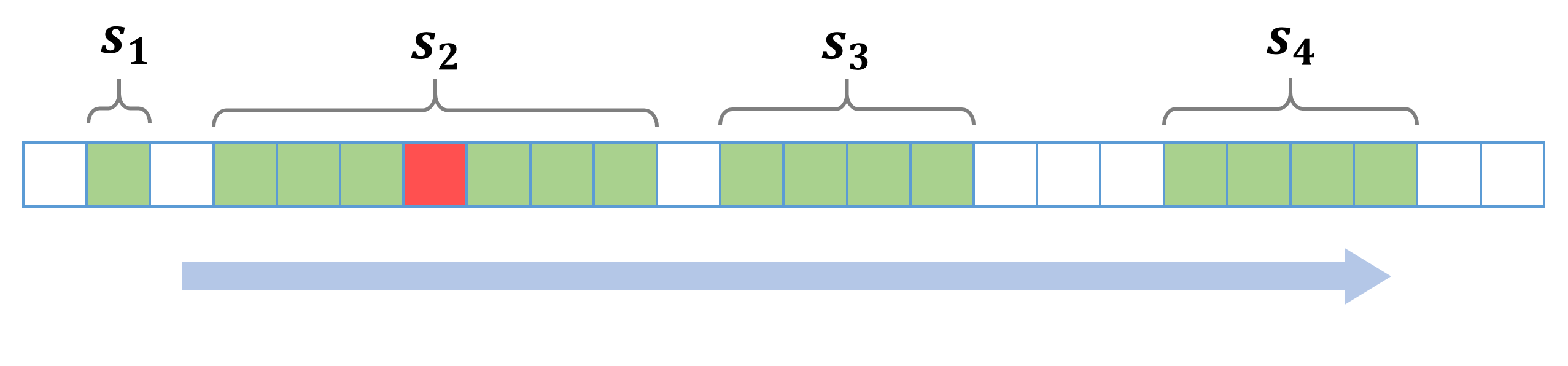} 
		
	\end{center} 
	\caption[Interconnected voxel sequences]{{\bf Interconnected voxel sequences.} Considering a linear scan from left to right (indicated by the lower horizontal blue arrow), four interconnected voxel sequences with scores greater than 0.95 are shown (green boxes), indicated by $s_1$, $s_2$, $s_3$ and $s_4$. The longest sequence corresponded to $s_2$. If this was the longest of all the scans, then the voxel in the middle of it (red box) was chosen as the approximate geometric center of the highest concentration of voxels.}
	\label{seq}
\end{figure}

It was assumed that all the voxels that correctly formed part of the whole tumor were connected to each other, that is, that each voxel with a TC label had as its immediate neighbor another voxel with a TC label. To exclude from the segmentation the voxels that were found far away and isolated, the following was done. A starting voxel located approximately at the geometric center of the highest concentration of TC-labeled voxels was chosen.
To find this voxel, the \emph{center} technique was used. Considering only the voxels that had a score greater than 0.95, linear scans were performed in the three spatial directions of the volume, recording the sequences of interconnected TC-labelled voxels (Fig \ref{seq}). Finally, the longest sequence of them was located, and the voxel located in the middle was chosen as the approximate geometric center of the region to be segmented.
Starting from the central voxel and identifying interconnected voxels within a score threshold, different segmentations of the whole tumor were obtained. However, by using different thresholds, the process of identifying interconnected voxels starting from the same central voxel involved a long computation time.
To resolve and improve this, a method called \emph{connected neighbors} was implemented.

\subsubsection{Connected neighbors}

\begin{figure}[ht] 
 	
 	\begin{center} 
 		\includegraphics[scale=0.17]{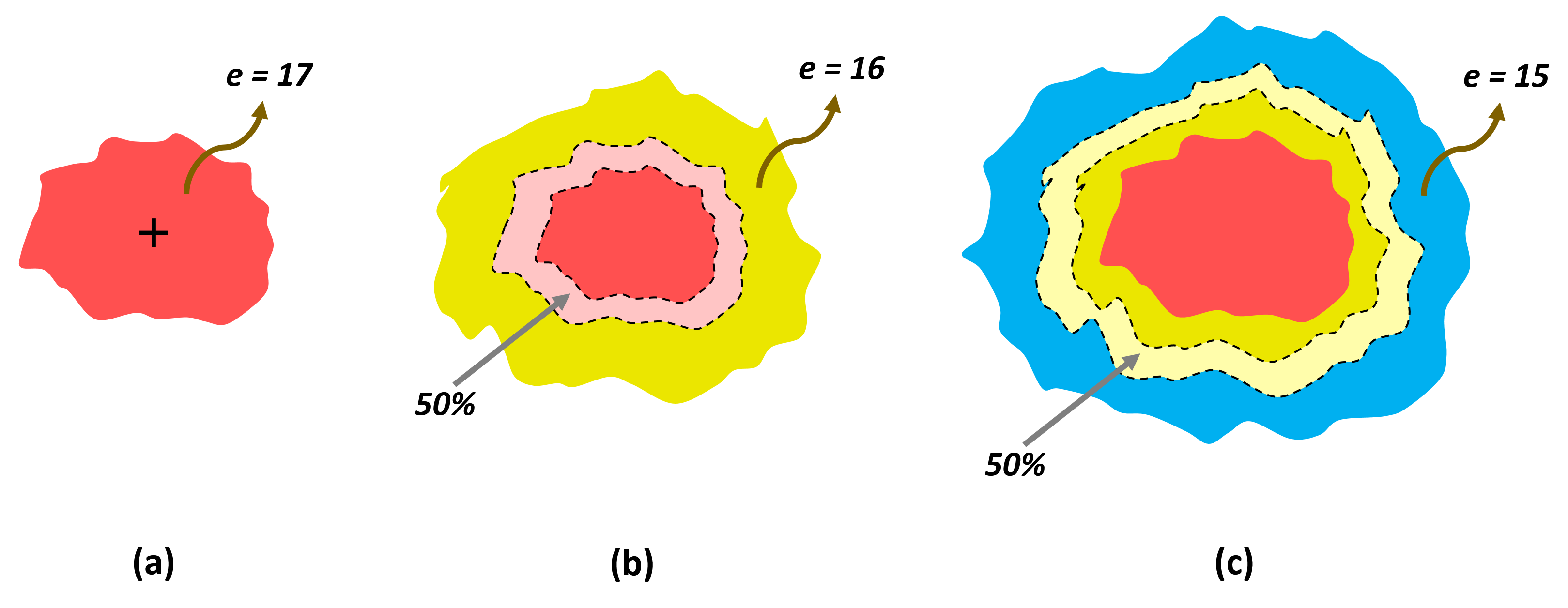} 
 		
 	\end{center}
 	\caption[{Connected Neighbors} to improve segmentation efficiency for different thresholds]{{\bf{Connected Neighbors}}. (a) Starting from the voxel located at the approximate geometric center (indicated by a black cross), the segmentation of the voxels whose score was greater than the threshold of 0.95 is shown. These had a numerical label equal to $e=17$ (red region). (b) The 50\% of the last voxels included in the previous segmentation are indicated (light red region). The voxels connected to each other whose score was within the threshold equal to 0.9 and which were connected at the aforementioned 50\% are shown. They had a numerical label equal to $e=16$ (yellow region). (c) Similarly as before, the voxels whose numerical label was equal to $e=15$ (blue region) are shown.}
 	\label{graf_cap} 
 	
 \end{figure}

\begin{figure}[ht] 
	
	\begin{center} 
		\includegraphics[scale=0.17]{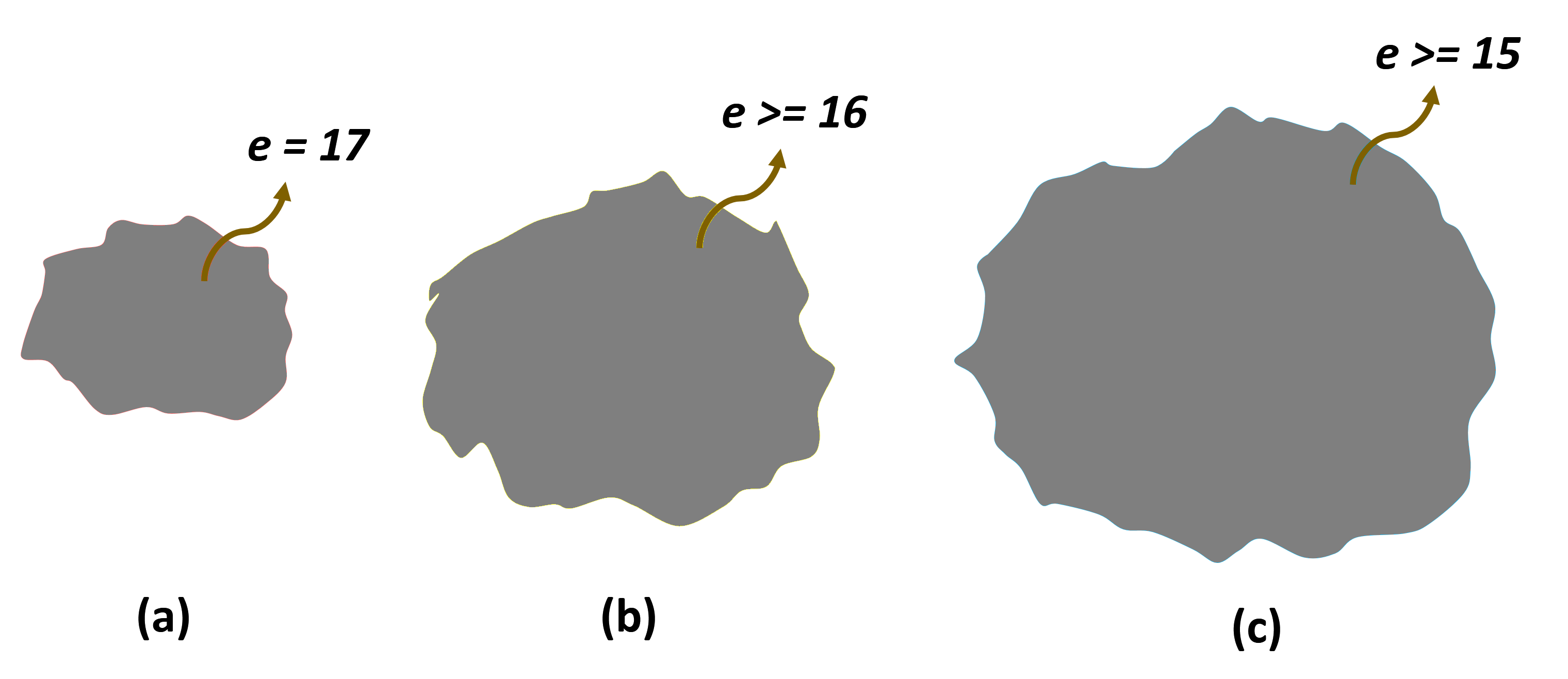} 
		
	\end{center} 
	\caption[Segmentations using numerical labels]{{\bf Segmentations using numerical labels.} From the volume of segmentations with numerical labels, some examples of segmentations of the whole tumor are shown using the threshold equal to 0.95, corresponding to the numerical label $e = $17(a); the threshold equal to 0.9, corresponding to labels greater than or equal to $e = 16$ (b); and the threshold equal to 0.85, corresponding to labels greater than or equal to $e = 15$ (c).}
	\label{seg_et}
\end{figure}

First, the segmentation of the whole tumor was obtained starting from the voxel located in the geometric center and identifying the interconnected voxels whose score was within the threshold equal to 0.95 (i.e., whose score was greater than or equal to 0.95). The identification of the voxels was done by locating the voxels one by one considering 16 possible directions in a three-dimensional space. Thresholds from 0.15 to 0.95 with steps of 0.05 were numbered from 1 to 17. Then, the voxels included in the aforementioned segmentation had the numerical label $e$ equal to 17 (Fig. \ref{graf_cap}(a)) . Afterwards, 50\% of the last voxels with numerical label equal to 17 added to the segmentation were located. Voxels connected to each other whose score was within the threshold equal to 0.9 and which were connected to 50\% of the voxels indicated above were located. This formed a cover of new voxels with numerical label equal to 16. Later and in a similar way, interconnected voxels whose score was within the threshold equal to 0.85 and that were connected to 50\% of the last voxels included with numerical label 16 were located. With this, another shell of new voxels whose numerical label was equal to 15 was added. This was repeated successively from threshold 0.95 to threshold 0.15 with steps of 0.05, assigning the numerical labels from 17 to 1. In the end, a volume with different shells associated with different numbers was obtained, en donde capas más internas contuvieron voxeles con puntuaciones más altas y capas más externas contuvieron voxeles con puntuaciones más bajas. Through this volume it was possible to obtain the segmentation of the whole tumor using any threshold, choosing the voxels whose numerical label was greater than or equal to the label of the respective threshold (Fig. \ref{seg_et}).

\begin{figure}[ht] 
	
	\begin{center} 
		\includegraphics[scale=0.18]{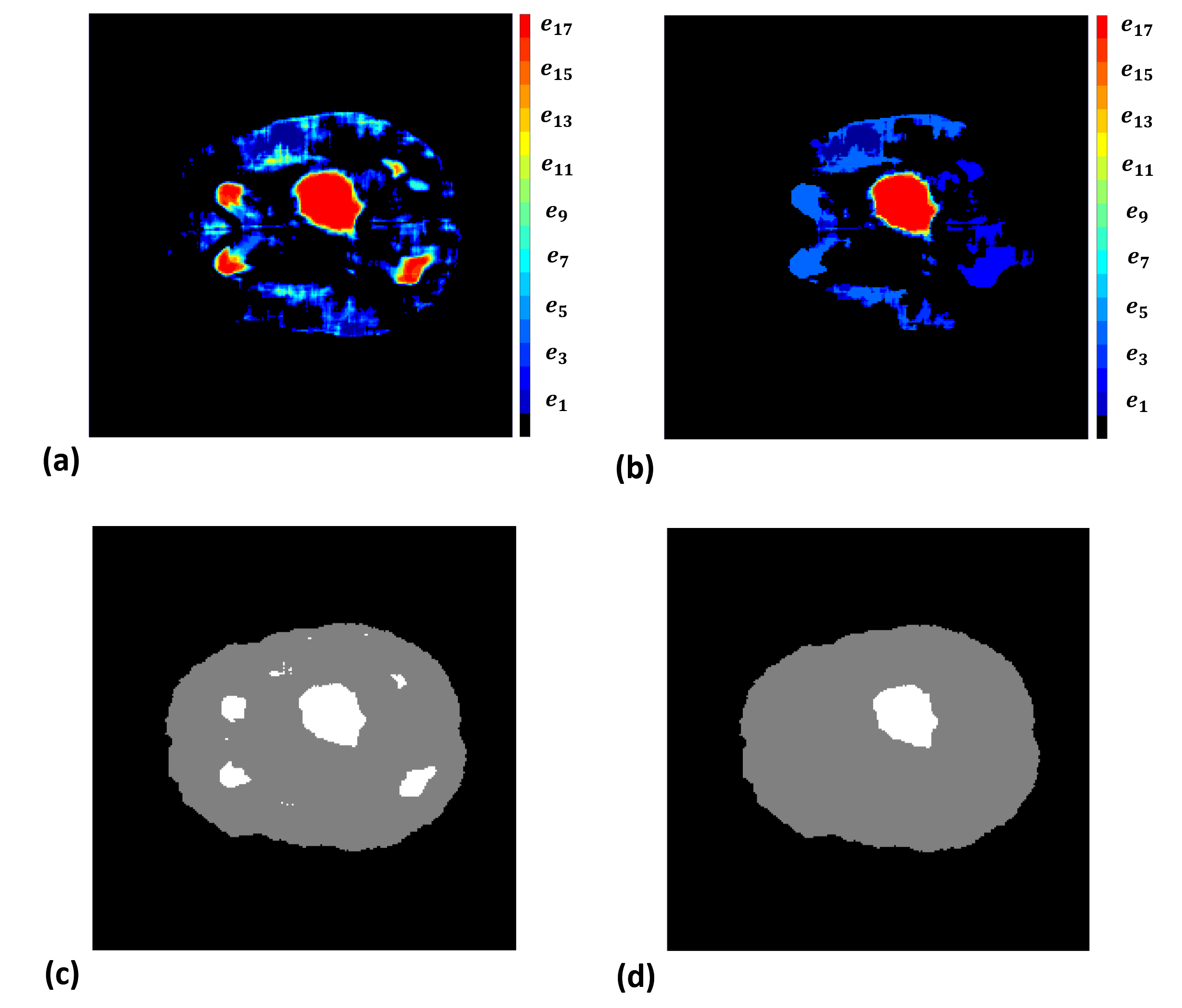} 
		
	\end{center}
	\caption[Example of exclusion of distant and connected voxels]{{\bf Example of exclusion of distant and connected voxels.} (a) Assignment of numerical labels to interconnected voxels starting from the central voxel without using the \emph{connected neighbors} technique. (b) Assignment of numerical labels using the \emph{connected neighbors} technique. Comparing with (a), it can be seen that distant voxels had lower numerical labels. (c) Whole tumor segmentation using some threshold from the labels in (a). (d) Whole tumor segmentation using the same threshold as in (c) but with the labels from (b).}
	\label{ej_excl} 
	
\end{figure}

\begin{figure}[ht] 
	
	\begin{center} 
		\includegraphics[scale=0.18]{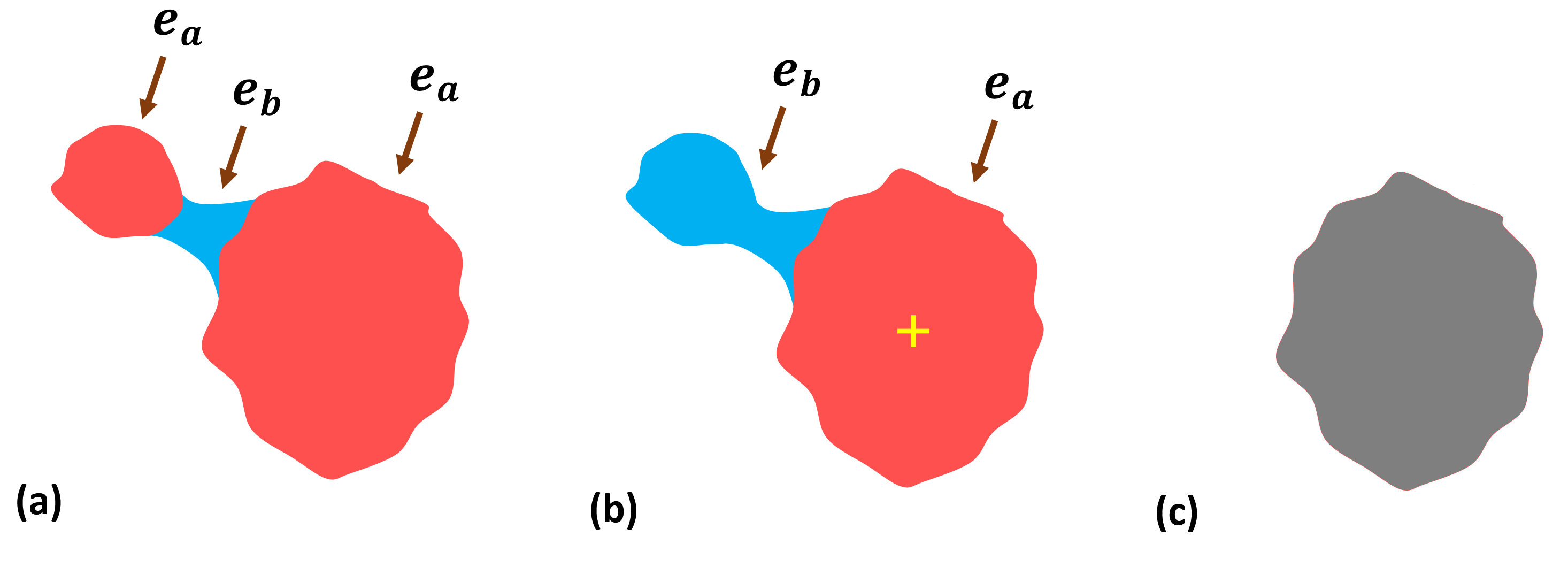} 
	
	\end{center} 
	\caption[Change from numeric label to separate voxel region]{{\bf Change from numeric label to separate voxel region.} (a) Two separate voxel regions with the same numeric label $e_a$ and connected through a third region with numerical label $e_b$, such that $e_a > e_b$. (b) Assuming that the segmentation starts from the central voxel indicated by a yellow cross, then only the voxels of that region kept the label $e_a$, while the second region changed its label to $e_a$. (c) Choosing an appropriate threshold (in this case $u_a$, associated to the numerical label $e_a$) the second region was excluded from the segmentation.}
		\label{si_conec}
\end{figure}

An example where the connected neighbors technique was not used is shown in Fig. \ref{ej_excl} (a). In this, a region composed of all the interconnected voxels was segmented starting from the central voxel. Afterward, each voxel was assigned a numerical label based on its score. As can be seen, there are many distant voxels that are connected to the highest concentration of them and that also have a large numerical label, so that, regardless of the threshold used, they would be erroneously included in the segmentation of the entire tumor (Fig. \ref{ej_excl}(c)). However, when using the connected neighbors technique, there was the advantage that far away voxels had a lower label (Fig. \ref{ej_excl}(b)) and by choosing an appropriate threshold they were correctly excluded from the whole tumor segmentation (Fig. \ref{ej_excl}(d)). To understand this, consider the following. Suppose that there were two regions made up of voxels with the same high numerical label and that they were connected through another region of voxels with a lower label (Fig. \ref{si_conec}). If the described method started in one of the first two mentioned regions, this region kept its higher numerical label while the other changed its label to a lower one. This occurred because the voxels in the second region were later considered when searching for interconnected voxels which had a lower label. Thus, the label of the remote voxels was changed and they were no longer part of the segmentation of the whole tumor.

Once the above was done, for a volume, the Dice coefficient was calculated between the real segmentation and each of the segmentations made using the thresholds that varied from 0.15 to 0.95 with steps equal to 0.05. This was repeated with all training and validation volumes. In the end, the threshold was chosen with which the validation volumes obtained, on average, the highest Dice coefficient. This threshold was called $u_p$.
Three other post-processing techniques were performed on the segmentations of the whole tumor considering the threshold $u_p$, which were called: \emph{refinement of TC}, \emph{fill by directions} and \emph{filtered}.

\subsubsection{Refinement of TC}

\begin{figure}[ht] 
	
	\begin{center} 
		\includegraphics[scale=0.35]{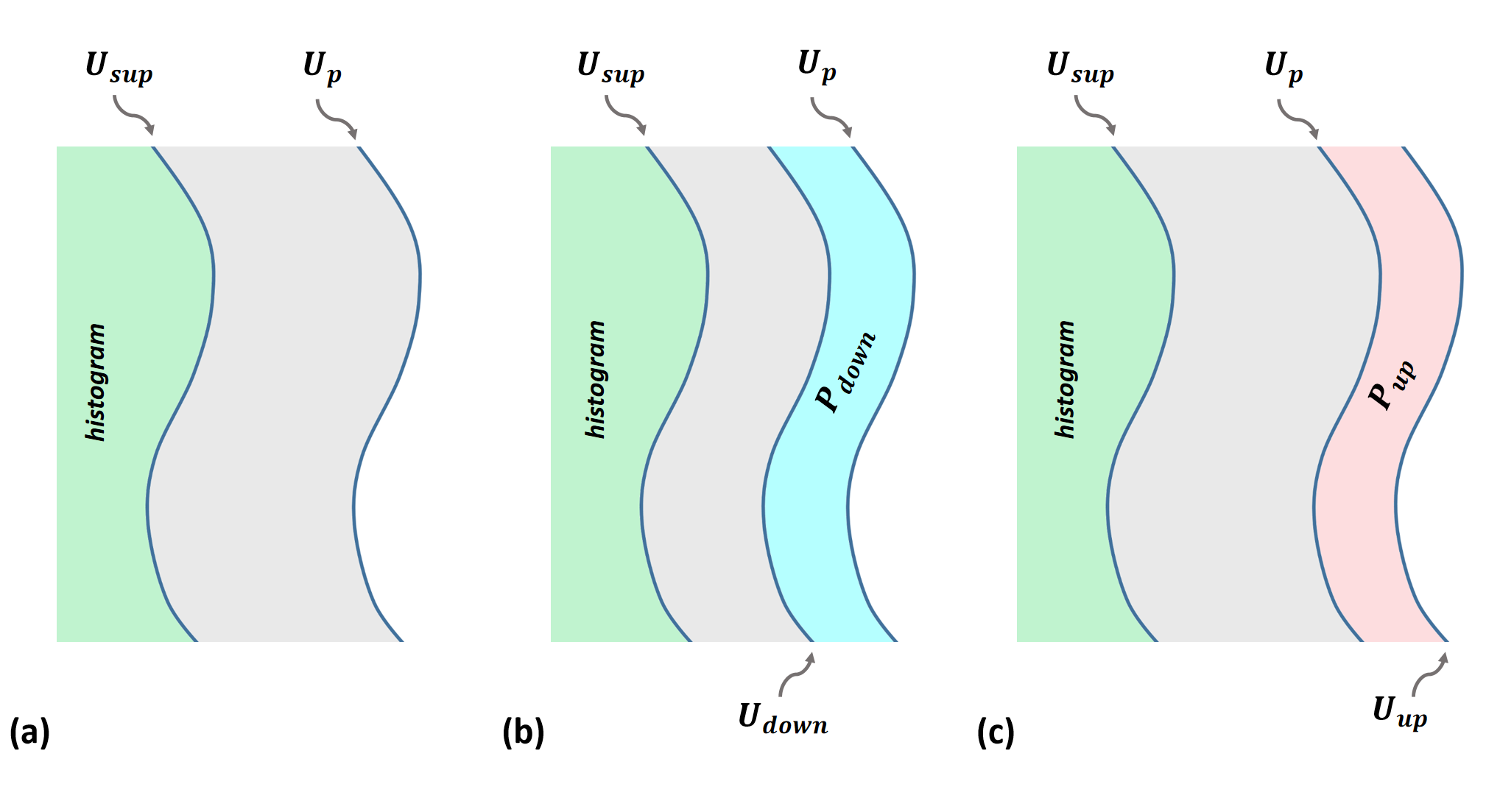} 
	
	\end{center} 
	\caption[Refinement of TC]{{\bf Refinement of TC. }(a) The initial segmentation made using $U_p$ (grey region), and the chosen region using $U_{sup}$ (green region) to compute the histogram are shown. (b) The regions ``below'' the initial segmentation located between $U_{down}$ and $U_p$ and (c) ``above'' located between $U_p$ and $U_{up}$ are indicated. From the first, the voxels whose intensities were outside the $P_{down}$ percentage were removed, and from the second, the voxels whose intensities were within the $P_{up}$ percentage were added.}
		\label{refine}
\end{figure}

The basic idea of \emph{refinement of TC} technique was to remove voxels in layers "below" the performed segmentation and add voxels "above" the same segmentation. The criteria to do this were based on the search for higher frequency intensities measured in the FLAIR modality, considering only voxels belonging to the segmentation that had higher scores. It was assumed that these higher frequency intensities were more representative of the whole tumor. To better understand this, consider the following.
Thresholds from 0.15 to 0.95 with steps of 0.05 were numbered from 1 to 17. The threshold number $u_p$, which was used to perform the segmentation, was called $U_p$. Another threshold was chosen whose number was called $U_{sup}$, such that $U_{sup} > U_p$. Voxels whose threshold number was within $U_{sup}$ had higher scores (Fig. \ref{refine}(a)). Considering only these voxels, a histogram of their intensity values measured in FLAIR was made. The histogram was calculated taking 100 intensity intervals ordered and of equal size considering all the intensities of the voxels. Another threshold was chosen whose number was called $U_{down}$, such that $U_{sup} > U_{down} > U_{p}$ (Fig. \ref{refine}(b)). A region of voxels "below" the initial segmentation and between the layers indicated by $U_{down}$ and $U_{p}$ was then obtained. Another threshold was chosen whose number was called $U_{up}$, such that $U_{sup} > U_{p} > U_{up}$ (Fig. \ref{refine}(b)). A region of voxels "above" the initial segmentation and between the layers indicated by $U_{p}$ and $U_{up}$ was then obtained (Fig. \ref{refine}(c)). It should not be forgotten that higher thresholds have higher associated numbers, and that these correspond to voxels with higher scores located in the innermost regions of the segmentation. Subsequently, two percentages were chosen, which were called $P_{down}$ and $P_{up}$. These represented the percentages of the most frequent intensities measured in the aforementioned histogram. Remembering that the histograms were measured in 100 equal intervals, then, if for example a percentage is equal to 40\%, then the 40 intensity intervals with the highest frequency were considered. The percentage $P_{down}$ was associated with the region of voxels "below" the initial segmentation and between the shells indicated by $U_{down}$ and $U_{p}$ (Fig. \ref{refine} (b)). The percentage $P_{down}$ was associated with the region of voxels "above" the initial segmentation and between the layers indicated by $U_{p}$ and $U_{up}$ (Fig. \ref{refine} (c)). Finally, to apply the refinement technique, the following was done. Voxels in the "below" region whose intensities were outside the percentage $P_{down}$ of most frequent intensities were removed from the initial segmentation. On the other hand, voxels in the "above" region whose intensities were within the $P_{up}$ percentage of the most frequent intensities were added to the initial segmentation. The specific values of alll these variables were chosen such that they improved the results obtained with the validation volumes.

\subsubsection{Fill by directions}

\begin{figure}[ht]
 	
 	\begin{center} 
 		\includegraphics[scale=0.18]{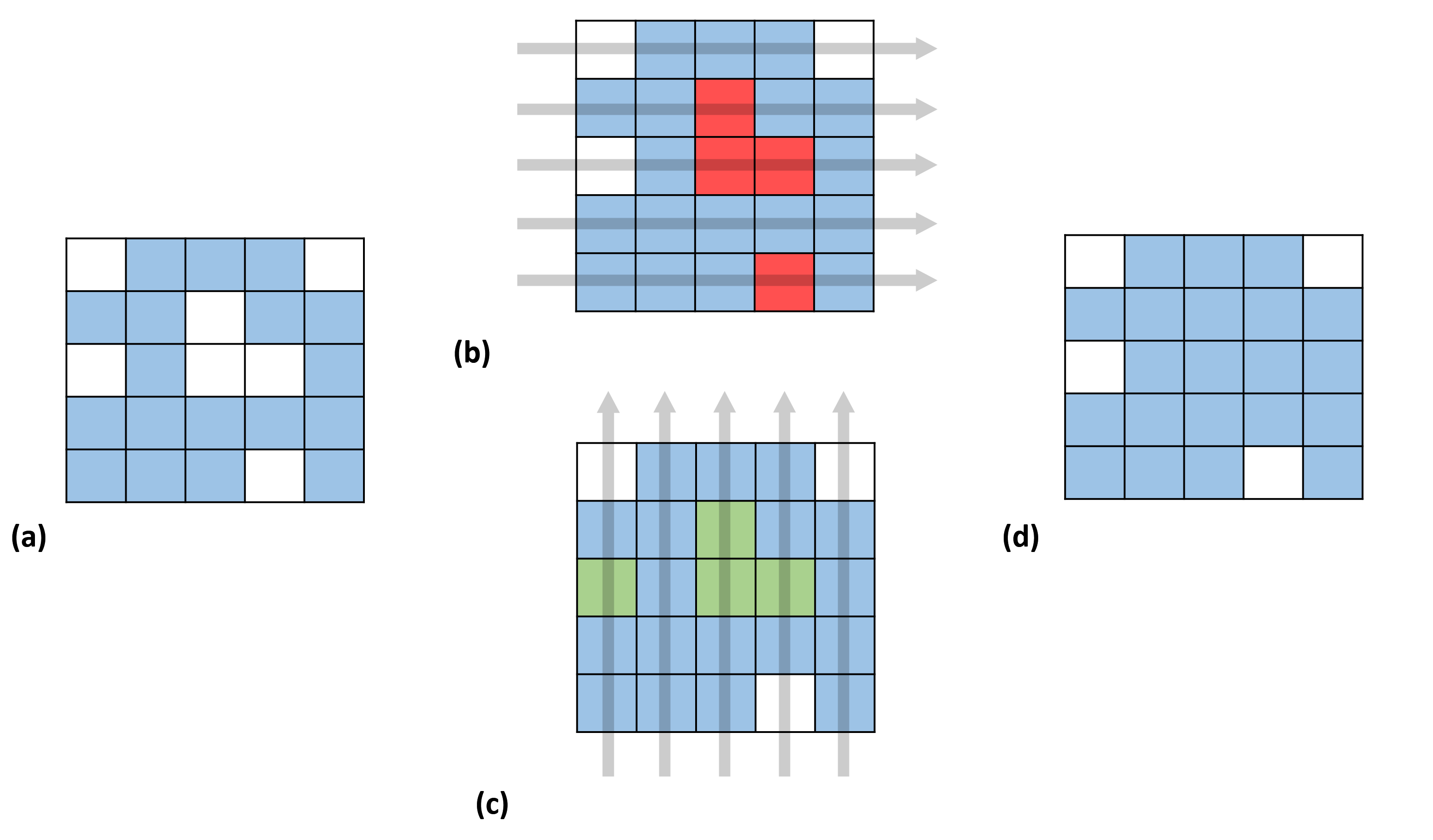} 
 		
 	\end{center}
 	\caption[Fill by directions]{{\bf Fill by directions.} (a) Segmented region (blue squares labeled TC) where a hollow is observed (three white squares labeled TS in the center). (b) Linear sweeps in one (horizontal) direction. Some voxels changed their label from TS to TC (red squares). (c) Linear sweeps in another spatial direction (vertical). Some voxels changed their label from TS to TC (green squares). (d) Only voxels that changed their label in all spatial directions (two in this example) kept the changed TC label.}
 	\label{relleno} 
 	
 \end{figure}

Hollows were usually observed in the segmentation of the whole tumor. These were regions composed of voxels labeled TS surrounded by voxels labeled TC (Fig. \ref{relleno}(a))
  Under the assumption that hollows shouldn't exist, it was then decided to ``fill'' them (i.e., change TS to TC). The \emph{fill by directions} technique consisted of the following. Linear scans were performed in the three spatial directions of the volumes. In each scan, the first and last voxel with a TC label were located. Between these two, all voxels that had TS labels were identified and their label changed to TC (Fig. \ref{relleno}(b) and (c)). Only the voxels that changed their label in all three scans of the three spatial directions kept their changed TC label (Fig. \ref{relleno}(d)). 

\subsubsection{Filtered}

\begin{figure}
	
	\begin{center} 
		\includegraphics[scale=0.5]{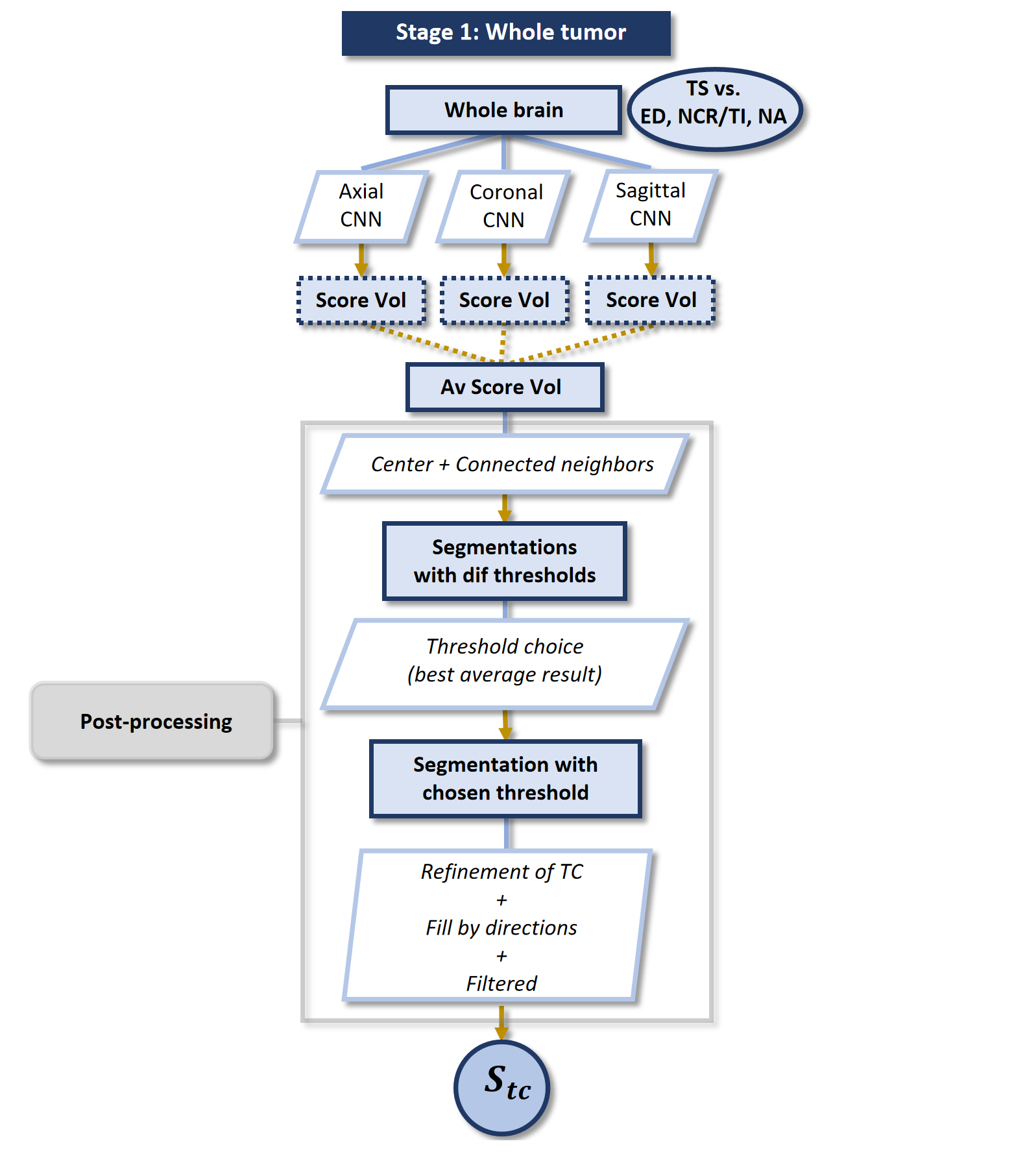} 
		
	\end{center} 
	\caption[Summary of stage 1]{{\bf Summary of stage 1.} The goal was to segment the whole tumor region, studying the whole brain and distinguishing between voxels labeled TS from voxels labeled ED, NCR/TI and NA. Three CNNs were applied to obtain three volumes of scores that were averaged to obtain a single volume of scores. Then the \emph{center} and \emph{connected neighbors} post-processing techniques were applied. A segmentation was obtained using a threshold with which the validation volumes obtained the best result on average. At the end, the techniques \emph{refinement of TC}, \emph{fill by directions} and \emph{filtering} were applied to obtain the volumes with the final segmentations of the whole tumor and which were called $S_{tc}$.}
	\label{resumen_1} 
\end{figure}

At this stage a median filter was applied. This filter calculated the median of the voxels that formed a cube of dimensions 3$\times$3$\times$3 and replaced the value of the voxel located in its center with the value of the median. Thus, this filter was applied to the entire volume after performing the \emph{fill by directions} technique. The volumes with the final segmentations of the complete tumor were called $S_{tc}$.
  These volumes were compared with the real segmentation by calculating the Dice coefficient. Fig. \ref{resumen_1} shows a diagram that summarizes all the procedures carried out in stage 1.

\subsection{Stage 2}

At this stage, only the voxels that formed the segmented whole tumor at stage 1 were considered.
A CNN was created for tumor core segmentation by differentiating ED labeled voxels (peritumoral edema) and NA or NCR/TI labeled voxels (which formed the tumor core). A new label called NT associated with the tumor core was defined. Of the total number of patches chosen to train the network, half had the label ED and the other half NT (Table \ref{parches_etapa2}), chosen randomly from all the possible patches of interest obtained from the 175 training volumes.
The CNN architecture of stage 2 was identical to the CNN architecture of stage 1 (Fig. \ref{red_1}). Only the optimizer used changed, being RMSprop for stage 1 and SGDM for stage 2. The output associated with the NT label was used and a threshold was chosen so that if the score of a voxel was greater than or equal to the threshold then the NT label was assigned, otherwise the ED label was assigned.

\begin{table}[ht]
	\centering
	
	\begin{tabular}{ccc}
		\hline 
		\textbf{Label}  & \textbf{Training patches}  &	  \textbf{Validation patches}     \\				
		\hline  
		ED & 250,000    &	50,000		\\	
		\hline 
		NA &	125,000   &	25,000		\\	
		\hline 
		NCR/TI &	125,000    &	25,000	\\
		\hline 
		Total & 500,000 & 100,000 \\
		\hline
	\end{tabular}
	\caption[Training and validation patches for stage 2]{{ \bf Training and validation patches for stage 2}. The total number of patches and their respective labels are indicated.}
	\label{parches_etapa2}
\end{table}

\subsubsection{Construction and combination of different CNNs}

\begin{figure}[ht] 
	
	\begin{center} 
		\includegraphics[scale=0.45]{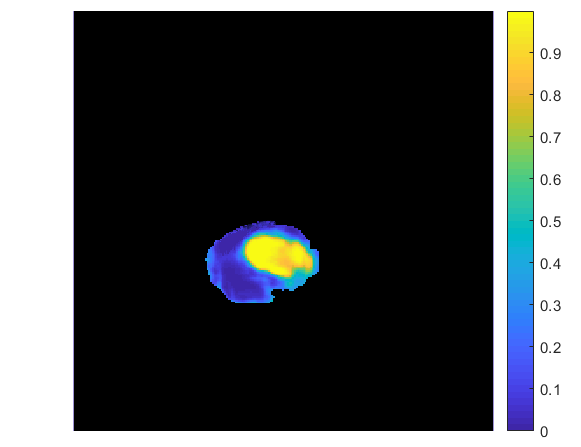} 
	
	\end{center} 
	\caption[Scores volume for stage 2]{{\bf Scores volume for stage 2.} The volume formed by the final scores of the voxels belonging to the segmented whole tumor in stage 1 is shown. The value of a score indicated the probability that a voxel corresponds to the NT label.}
		\label{vol_punt_2} 
\end{figure}

As before, in step 2 a CNN was created for each of the axial, coronal and sagittal planes. The same training voxels are studied using their patches seen from each plane. Three scores were then obtained for a voxel, which were averaged to obtain a single final score. A score volume was formed from the final scores of the voxels (Fig. \ref{vol_punt_2}).
Different thresholds were then used, such that if a voxel score was within the threshold then it was assigned the NT label. In total 17 thresholds were considered varying from 0.15 to 0.95 with steps of 0.05.
Tumor core segmentations were obtained after applying post-processing techniques called \emph{filtered} to the score volumes.  Subsequently, the technique called \emph{threshold customization with RNN} was applied to each volume to individually choose the threshold that would give the best segmentation of the tumor core.

\subsubsection{Filtered}

\begin{figure}[ht]
	
	\begin{center} 
		\includegraphics[scale=0.22]{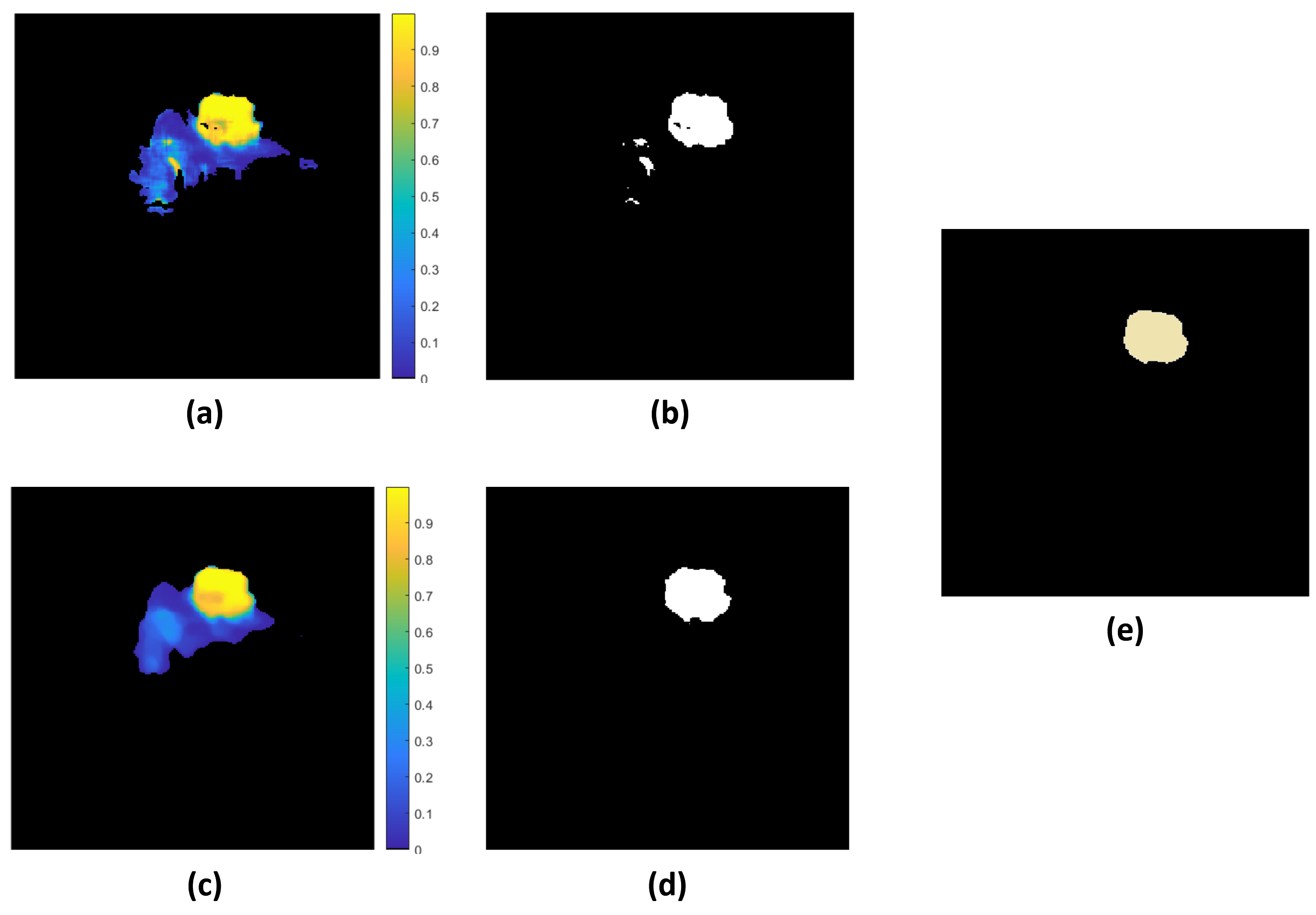} 
		
	\end{center} 
	\caption[Filtering to remove distant voxel clusters]{{ \bf Filtering to remove distant voxel clusters.} (a) Score volumes with small clusters formed by voxels with high scores (color close to yellow), surrounded by voxels with low scores ( color close to blue) and away from the highest concentration of them. (b) In this example, by choosing an appropriate threshold equal to 0.45, the segmented region that would correspond to the nucleus of the tumor is shown. (c) Score volume after applying the median filter several times. The scores of the small clusters of voxels indicated in (a) were reduced. (d) Using the same threshold of 0.45, the segmentation of the tumor core is shown, in which the small distant clusters are no longer included. (e) Actual segmentation.}
	\label{1_4_2_filt} 
\end{figure}

The median filter was applied to the score volumes a certain number of times. This not only smoothed out the volumes. After choosing an appropriate threshold, the filter also removed small clusters of voxels with high scores that were far from the segmentation of interest (when they existed) and that were erroneously labeled as TC (Fig. \ref{1_4_2_filt}).
This occurred because the median filter reduced the value of scores for voxels that were surrounded by other voxels with lower scores. This was the case for the aforementioned small voxel clusters. Then, by choosing an appropriate threshold, these voxels were excluded from the core tumor segmentation.
The number of times the filter was applied depended on the average improvement seen in the validation volumes. To choose the appropriate threshold, a post-processing called \emph{customization of the threshold with RNN} was performed.

\subsubsection{Threshold customization with RNN}

\begin{figure}[ht] 
	
	\begin{center} 
		\includegraphics[scale=0.3]{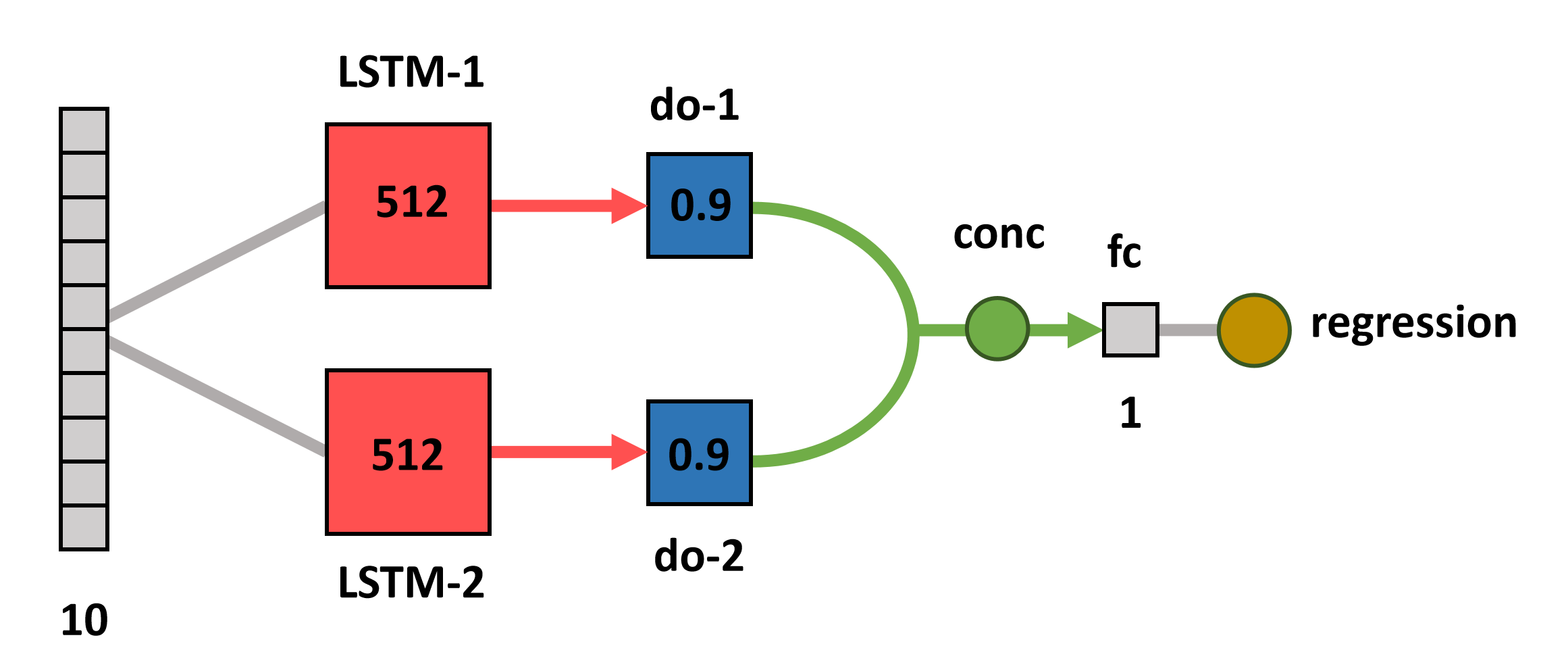} 
	\end{center} 
	\caption[RNN architecture for threshold customization]{{\bf RNN architecture for threshold customization.} The input sequence of 10 ordered values is shown, followed by two LSTMs (LSTM-1 and LSTM-2) with 512 hidden units each. A dropout of 0.9 was applied to the LSTM starts. The result of both processes was concatenated and fully connected to a single node layer. At the end, a regression was performed and a single output equal to the custom threshold was obtained. }
	\label{RNR} 
\end{figure}

To make the individual choice of the most appropriate threshold for each volume, a recurrent neural network (RNN) was built. This customization only made sense if better results were obtained through the RNN compared to choosing the same threshold for all validation volumes. The RNN entries were obtained in the following way. For each volume, a histogram was calculated from its score volume without filtering. Histogram intervals were different for each. Considering a particular volume, first the maximum value $p_{max}$ of the scores was obtained, which would be the upper bound of the intervals. Then the lower bound of the intervals was calculated as $p_{inf} = p_{max} - $ 0.01. Finally, the intervals varied from $p_{inf}$ to $p_{max}$ with steps of 0.001. Thus, the RNN entries were sequences of 10 ordered values corresponding to the frequencies of the 10 intervals of highest scores. For a volume, its expected output from the network was the appropriate threshold with which it obtained the best segmentation of the tumor core. To know this, the segmentations were previously performed using all the thresholds and the most appropriate threshold was identified. The inputs that were used to train the network were the sequences of the 175 training volumes. During training, the network was validated with the corresponding sequences from the 39 validation volumes. The architecture of the RNN is shown in Fig. \ref{RNR}. The batch size was equal to 175, this being the total number of training entries. The network was trained on 10,000 epochs. The RMSprop optimizer was used with a learning range of 10$^{-4}$.

Once the thresholds were customized and the segmentations were obtained for each volume, the results were compared with those obtained by using the same threshold $u^*$ for all the volumes, with which the best results were obtained on average. Assuming that the results improved after customizing the thresholds, then another post-processing technique called \emph{fill by planes} was applied, similar to the one used in stage 1. In case better results were not achieved with the threshold customization , this technique was skipped and continued with only the \emph{fill by planes}.

\subsubsection{Fill by planes}

\begin{figure} 
	
	\begin{center} 
		\includegraphics[scale=0.55]{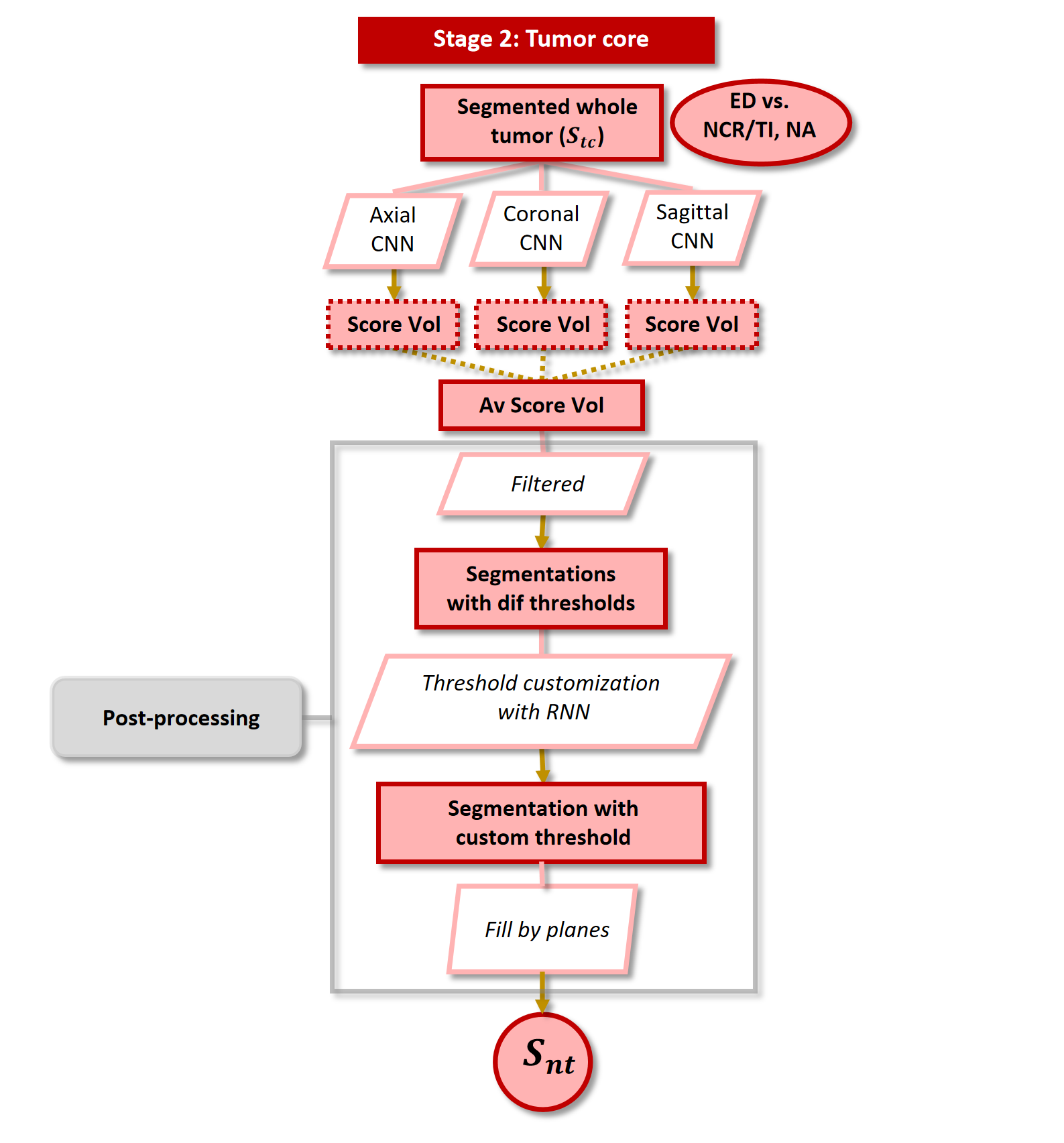} 
	\end{center} 
	\caption[Summary of stage 2]{{\bf Summary of stage 2.} The goal was to segment the tumor core by studying only the region of the whole tumor that was segmented in stage 1, distinguishing between ED labeled voxels and NCR/TI and NA labeled voxels. Three CNNs were created, each from the axial, coronal, and sagittal planes of the MRI modalities. Score volumes were obtained and averaged to obtain a single score volume. This volume was \emph{filtered} and different segmentations were obtained using different thresholds. For each volume, the post-processing technique called \emph{customization of the threshold with RNN} was applied to find the appropriate threshold to obtain the best segmentation of the core tumor. At the end, the \emph{fill by planes} technique was applied and the segmentation of the tumor core called $S_{nt}$ was obtained.}
	\label{resumen_2} 
\end{figure}

Following a similar idea to the \emph{fill by directions} technique of stage 1, in the \emph{fill by planes} of stage 2, scans were made in the axial, coronal and sagittal planes. The scans in each plane were equivalent to performing two linear scans. For example, considering the three spatial directions $x$, $y$ and $z$, and the planes $xy$, $yz$ and $xz$, then the scans in the $xy$ plane were equal to two linear scans, one in the $x$ direction and one in the $y$ direction. In each linear scan, ED labeled voxels that were in the middle of NT labeled voxels changed their label to NT. In each plane the voxels kept the change if it occurred in the two linear scans. Afterwards, of these voxels only those that changed their labels in the three planes kept the change at the end.
The segmentation of the tumor core was named $S_{nt}$. These volumes were compared to the actual segmentations using Dice's coefficient. Fig. \ref{resumen_2} shows the summary of the processes applied in stage 2.

\subsection{Stage 3}

The goal of stage 3 was to segment the enhancing tumor core whose voxels were labeled NA.
 This stage involved the segmentation of the most difficult region of the tumor. One of the reasons for its complexity was that in some cases the size of the region to be segmented was smaller compared to the two previous stages.
 Also, although the size of the region was not relatively small, at other times it was made up of many small regions scattered but close to each other.
In addition, the segmentation stages were performed sequentially, that is, the study region was restricted to that segmented in a previous stage. Consequently, stage 3 preserved the errors made in the previous two stages. Therefore, at this stage it was not convenient to be restricted to the region segmented in stage 2. Thus, in three different parts of this stage, the segmentations of the whole tumor and the tumor core, obtained in stages 1 and 2 respectively, were studied. besides the whole brain.
The creation of three CNNs was proposed, which studied different regions and labels. Subsequently, two post-processing techniques called \emph{filtered} and \emph{refinement by layers} (different from the previous stages) were applied. On the other hand, not all tumors had an enhancing tumor core, so there was no region to segment. However, because no neural network was perfect, almost anyone could assign the NA label to a certain number of voxels in volumes where no voxels with this label actually existed. In these cases, when calculating the Dice coefficient between the erroneous segmentation and the real segmentation, the result was equal to zero. Therefore, the post-processing technique called \emph{non-existence of NA} was proposed. In this technique, a model was created to predict whether or not the volume to be segmented contained this region. Thus, when the model predicted that the region did not exist, then the segmentation that had been made of it was eliminated. And in case the model predicted that it existed, then the segmentation made was preserved.
Later, at this stage, a post-processing technique was also applied to choose the appropriate threshold with which a volume would obtain the best segmentation of the enhancing tumor core. This technique was called \emph{threshold customization with regression}. Finally another technique called \emph{refinement of NA} was applied.

\subsubsection{Creation of three CNNs}

In general, the three proposed CNNs assigned one of two possible labels to the studied voxels.
Table \ref{etiq_et3} shows the regions studied by the CNNs and their respective labels. Table \ref{tab_parch} shows the number of training and validation patches used for each label.

\begin{table}[ht]
	\centering
	
	\begin{tabular}{cccc}
		\hline 
		\textbf{No. CNN}  & \textbf{Region studied}  & \textbf{Label 1} & \textbf{Label 2}    \\				
		\hline  
		1 & Tumor core    &	NA	& NCR/TI	\\	
		\hline 
		2 &	Whole tumor   &	NA & NCR/TI + ED		\\	
		\hline 
		3 &	Whole brain    &	NA	 &  NCR/TI + ED + TS\\
		\hline 
	
	\end{tabular}
	\caption[Regions and labels studied for each CNN]{{\bf Regions and labels studied for each CNN.} In the first two CNNs, the regions studied were those segmented in the previous two stages. In all the CNNs, one of two labels was assigned, the first being the NA label and the second one formed by the other labels involved in the region studied.  }
	\label{etiq_et3}
\end{table}

\begin{table}[ht]
	\centering
	
	\begin{tabular}{c|ccc}
		\hline 
		\textbf{No. de CNN}  & \textbf{Label}  & \textbf{Training patches}  &	  \textbf{Validation patches}     \\				
		\hline  
		 & NA & 250,000    &	50,000		\\	
		 
1	&	NCR/TI &		250,000     &	50,000	\\		
	
		&Total & 500,000 & 100,000 \\
		\hline
		 &NA & 250,000    &	50,000		\\	
	
	2	&NCR/TI &		125,000     &	25,000	\\	
	 
		&ED & 125,000 & 25,000 \\	
	 
		&Total & 500,000 & 100,000 \\
		\hline
		 & NA & 249,999    &	49,998		\\	
		
	&	NCR/TI &		83,333     &	16,666	\\	
		
	3	&ED & 83,333 & 16,666 \\	
		
		&TS & 83,333 & 16,666 \\	
		
		&Total & 499,998 & 99,996 \\
		\hline
	\end{tabular}
	\caption[Training and validation patches for stage 3]{{\bf Training and validation patches for stage 3 CNN 1}. The total number of patches studied and their respective labels are indicated for each CNN.}
	\label{tab_parch}
\end{table}

\begin{figure}[ht] 
	
	\begin{center} 
		\includegraphics[scale=0.4]{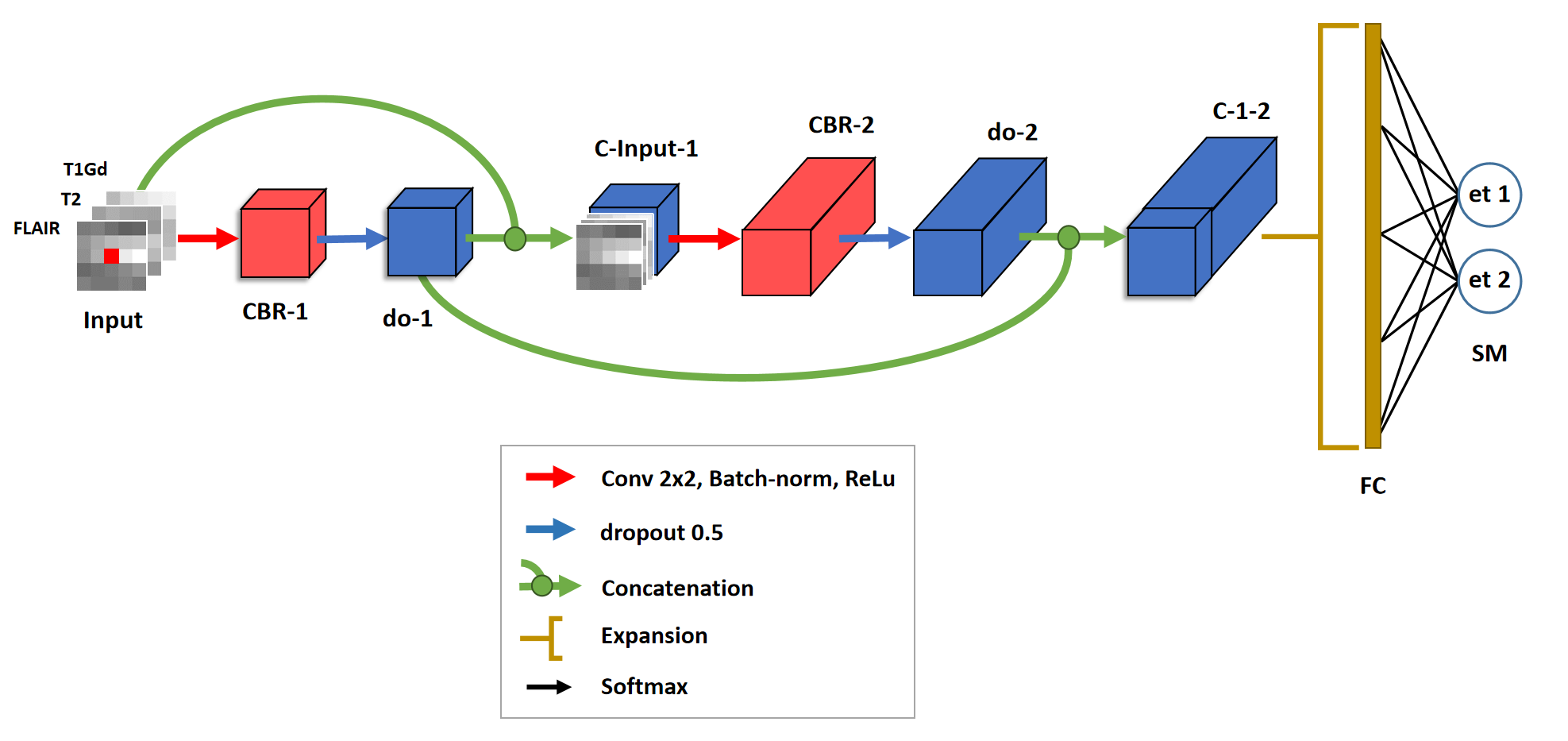} 
		
	\end{center} 
	\caption{{\bf Architecture for the three CNNs built in stage 3.} The network input is shown as a patch of 5$\times$5$\times$3 voxels. Only the voxel located in the center of the patch was labeled. At the end, the softmax (SM) function was applied to obtain two outputs related to the labels of interest (et 1 and et 2). }
	\label{red_3_1} 
\end{figure}

All training and validation patches with the labels of interest were randomly chosen from all possible patches obtained from the 175 training gliomas and 39 validation gliomas, respectively.
The same architecture shown in Fig. \ref{red_3_1} was used in all the CNNs, with the only difference being the patches and labels used.
Considering a volume plane (axial, sagittal or coronal), the network input (Input) was composed of three patches of dimensions 5$\times$5 from the same region, but seen in the MRI modalities T\textsubscript{1Gd }, T\textsubscript{2} and FLAIR. The voxel to be labeled was located in the center of the patch. A convolution layer with filter size equal to 2$\times$2, with stride equal to 1 and padding such that the output had the same dimensions as the input was applied. The number of filters was equal to 168. It was followed by a ReLu activation and batch normalization layer (CBR-1). Then a dropout of 0.5 (do-1) was made. Subsequently, a concatenation was made between the network input and the droput result (C-Input-1). Then, the above was repeated but using 256 convolution filters (CBR-2) and performing a concatenation between the outputs of the two dropouts (C-1-2). Finally, this volume was expanded as a column of dimensions 1$\times$1$\times$9600 nodes and the softmax function was applied to obtain two outputs related to the labels of interest. In the three CNNs, only the output associated with the NA label was considered.
Table \ref{tab_tam_vol} shows the output size of each layer.

\begin{table}[ht]
	\centering
	
	\begin{tabular}{cc}
		\hline 
		\textbf{Layer}  & \textbf{Output size}      \\				
		\hline  
		Input & 5$\times$5$\times$3  	\\	
		\hline 
		CBR-1 & 5$\times$5$\times$128	\\	
		\hline 
		do-1 & 5$\times$5$\times$128 \\	
		\hline 
		C-Input-1 & 5$\times$5$\times$171 \\	
		\hline 
		CBR-2 & 5$\times$5$\times$256 	\\	
		\hline 
		do-2 & 5$\times$5$\times$256 \\
		\hline 
		C-1-2 & 5$\times$5$\times$384	\\	
		\hline 
		FC & 1$\times$1$\times$9600	\\	
		\hline 
		SM & 1$\times$1$\times$2	\\	
		\hline 
	\end{tabular}
	\caption[Layers and their output sizes]{{\bf Layers and their output sizes.} Information about the size of the outputs of each of the layers of the CNN is indicated.}
	\label{tab_tam_vol}
\end{table}

For each of the three CNNs, a total of three networks were created by studying the same voxels but using the patches viewed from the axial, coronal, and sagittal planes, respectively. A score volume was obtained from each of these three networks, which were then averaged to obtain a single score volume. The averaged score volumes were called $V_{nt}$, $V_{tc}$ and $V_{tc}$, obtained from CNN 1, 2 and 3 respectively.

\subsubsection{Gaussian filter}

\begin{figure}[ht] 
	
	\begin{center} 
		\includegraphics[scale=0.2]{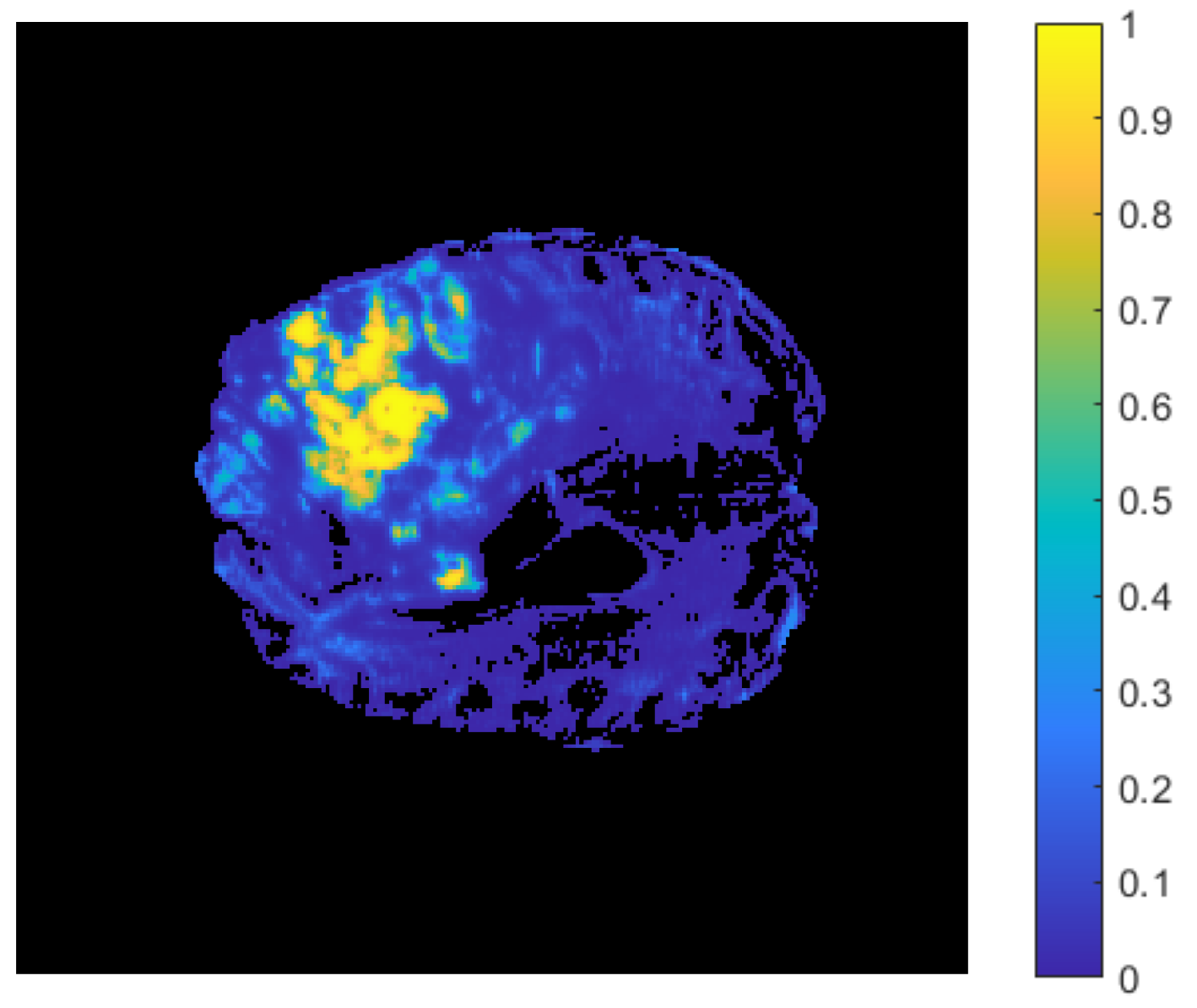} 
		
	\end{center} 
	\caption[Score volume for stage 3]{{\bf Score volume for stage 3.} Volume $V_{na}$ obtained by averaging the volumes $V_{nt}^f$, $V_{tc}^f$ and $V_ {cc}$.}
	\label{et3_prom} 
\end{figure}

Gaussian filters with different values of $\sigma$ were applied to the averaged score volumes $V_{nt}$ and $V_{tc}$. For the volume $V_{nt}$, a $\sigma =$ 0.5 with a filter size equal to 3$\times$3$\times$3 was used, obtaining the volume $V_{nt}^f$. For the volume $V_{tc}$, a $\sigma =$ 1.5 and a filter size of 7$\times$7$\times$7 were used, obtaining the volume $V_{tc}^f$. For the volume $V_{cc}$ the filter was not applied.
Afterwards, these filtered volumes were averaged and a single score volume $V_{na}$ was obtained  (Fig. \ref{et3_prom}).
Subsequently, the volume $V_{na}$ was applied a post-processing technique called \emph{refinement by layers}, while different thresholds were used to obtain the segmentation of the enhancing tumor core.

\subsubsection{Refinement by layers}

\begin{figure} 
	
	\begin{center} 
		\includegraphics[scale=0.2]{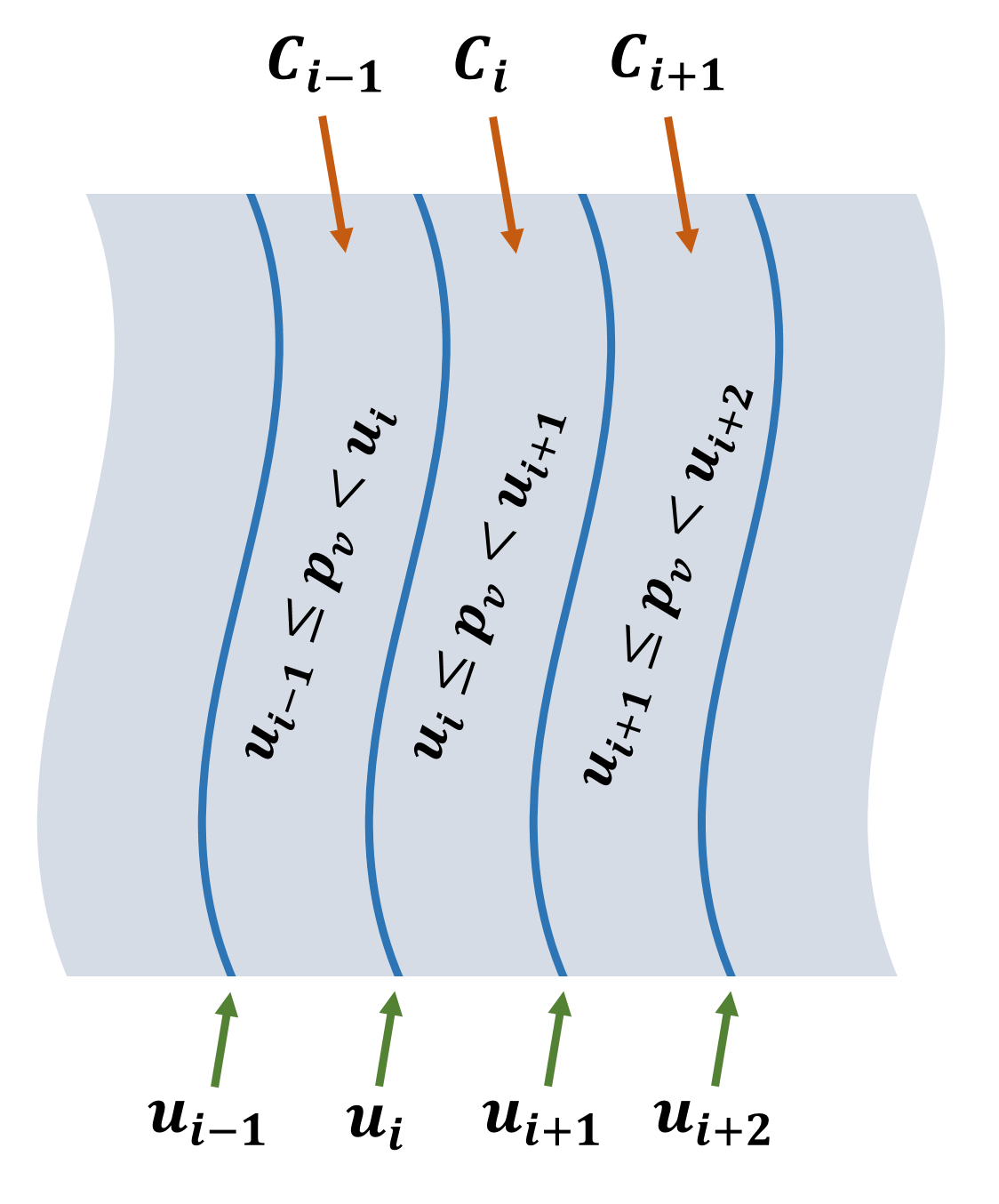} 
		
	\end{center} 
	\caption[Section of a score volume divided into uniform layers]{{\bf Section of a score volume divided into uniform layers.} Three intermediate layers are shown ($C_{i-1}$, $C_i$ and $C_{i+1 }$), indicating the condition met by the scores $p_v$ of the voxels contained in the layers and the threshold associated with their boundaries ($u_{i-1}$, $u_{i}$, $u_ {i+1}$ and $u_{i+2}$ respectively), with $i =$ 1, 2, 3,..., 16.} 
	\label{ref_cap_1} 
\end{figure}

\begin{figure}[ht] 
	
	\begin{center} 
		\includegraphics[scale=0.4]{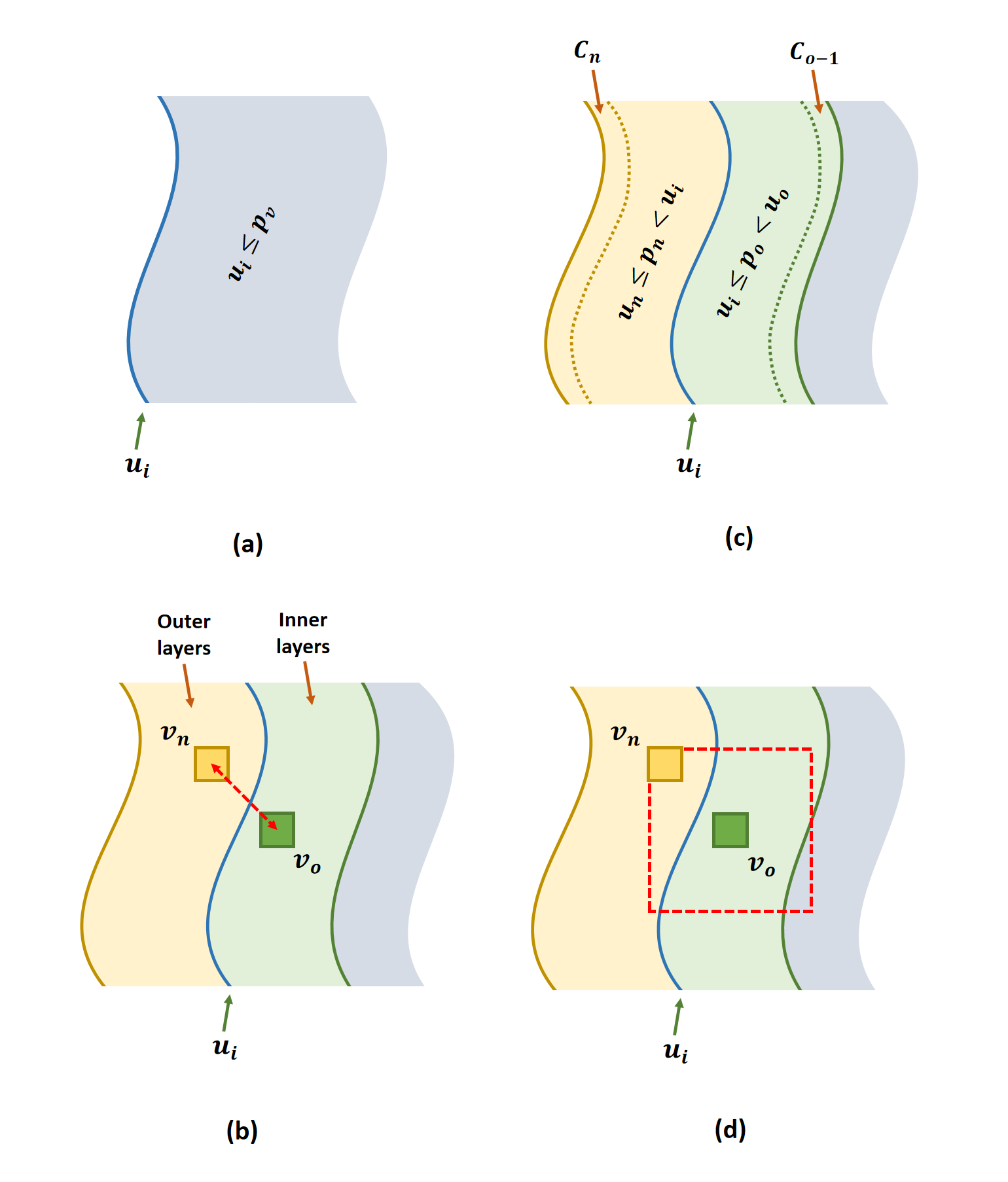} 
		
	\end{center} 
	\caption[Refinement by layers]{{\bf Refinement by layers.} (a) Initial segmentation (grey region) of the enhancing tumor core using the threshold $u_i$. (b) Inclusion of voxels $v_n$ located in outer layers separated a maximum distance from voxels $v_o$ located in inner layers. (c) Conditions that met the scores $p_n$ and $p_o$ of the voxels $v_n$ and $v_o$ respectively. (d) Implementation of cubes (shown as a red dotted square) with dimensions equal to the aforementioned maximum distance. In the center of each cube, a voxel $v_o$ was found. Only voxels $v_n$ located inside the cube were added to the segmentation.} 
	\label{all_cap} 
\end{figure}

Thresholds from 0.15 to 0.95 with steps of 0.05 were applied to the score volumes $V_{na}$ to obtain different segmentations of the enhancing tumor core. At the same time that a threshold was used, a technique called \emph{refinement by layers} was also applied. To simplify the explanation of the technique, it is convenient to assume that the score volume could be divided into uniform layers $C_{i}$, each one containing voxels whose score $p_v$ was such that $u_i \leq p_v < u_{i+1 } $, with $i = 1, 2, ..., 16$, and $u_i$ varying from 0.15 to 0.95 with steps of 0.05 (Fig. \ref{ref_cap_1}). A final layer $C_{17}$ contained voxels whose score $p_v$ was such that $u_{17} \leq p_v$, where $u_{17} =$ 0.95. Suppose that a segmentation was obtained after applying the threshold $u_i$, so that it contained all the voxels whose score $p_v$ was such that $u_i \leq p_v$ (Fig. \ref{all_cap}(a)). So, the basic idea of \emph{refinement by layers} was to add to the segmentation voxels $v_n$ located inside one or several external layers, being further away at a certain maximum distance from other voxels $v_o$ contained in one or several layers internal (Fig. \ref{all_cap}(b)). In other words, the scores $p_n$ of the voxels $v_n$ were such that $u_n \leq p_n < u_{i}$, being $u_{n}$ the threshold associated to the last external layer $C_{n }$ considered; and the scores $p_o$ of the voxels $v_o$ were such that $u_i \leq p_o < u_o$, being $u_o$ the threshold associated to the first internal layer $C_{o-1}$ considered (Fig. \ref {all_cap}(c)). On the other hand, regarding the maximum distance between the voxels $v_n$ and $v_o$, this idea was implemented by choosing cubes of dimensions $d\times d \times d$ that surrounded the voxels $v_o$ and it was established that the voxels $v_n$ had to be inside the cubes (Fig. \ref{all_cap}(d)). The values of number of inner and outer layers, in addition to the size of the cube mentioned, were chosen such that the best possible results were obtained with the validation volumes. In order to do the latter, the appropriate threshold that allowed obtaining the best segmentation of the enhancing tumor core was chosen for each volume. This was done through a threshold customization technique. However, before that, the technique called \emph{non-existence of NA} was applied.

\subsubsection{Non-existence of NA}

The \emph{non-existence of NA} post-processing technique studied some variables extracted from the score volumes, to identify whether or not they actually contained voxels with the NA label.
The variables studied related two quantities. One of them was the number of voxels $N_{nt}$ labeled as NT (those that formed the segmented core tumor in stage 2). The other was the number of voxels $N_{na}^j$ labeled as NA (those that formed the segmented enhancing tumor core at stage 3). The superscript $j$ indicated the threshold number used. Then, the relationships between these quantities were as follows:

	\begin{eqnarray}
	\label{rel}
	P^j =  \dfrac{ N_{nt} }{ N_{na}^j },  \ \ \ \ \  \ \ \ \ \ Q^j = \dfrac{N_{na}^j}{N_{nt}},  \ \ \ \ \  \ \ \ \ \ R^j = \dfrac{N_{na}^j}{N_{na}^j + N_{nt}}.
	\end{eqnarray}


If during the segmentation of the enhancing tumor core using some threshold $u_j$ no voxel was assigned the NA label, then the amount $N_{na}^j$ was equal to zero. In that case, the following was obtained:

\begin{eqnarray}
\label{rel_0}
P^j \rightarrow  \infty,  \ \ \ \ \  \ \ \ \ \ Q^j = 0,  \ \ \ \ \  \ \ \ \ \ R^j = 0.
\end{eqnarray} 


For convenience and arbitrarily, it was decided to make $P^j$ equal to 10, thus having:

\begin{eqnarray}
\label{rel_01}
P^j =  10,  \ \ \ \ \  \ \ \ \ \ Q^j = 0,  \ \ \ \ \  \ \ \ \ \ R^j = 0.
\end{eqnarray}

This occurred in one of two possible cases. The first was when the threshold $u_j$ used was large enough that no voxel reached the necessary score to be labeled as NA. The second case was when, regardless of the threshold used, no voxel was assigned the NA label (ideal situation for volumes that did not really have the enhancing tumor core). Due to the first case, it was decided to study only the quantities using the thresholds from 0.15 to 0.85 with steps of 0.05, thus excluding the thresholds 0.9 and 0.95.
The reason for studying the quantities $N_{nt}$ and $N_{na}^j$, in addition to the relationships in eqs. \ref{rel_01}, was the following. It was observed that usually when the volumes did not really have the enhancing tumor core, a relatively small number of voxels were labeled as NA compared to the total number of voxels labeled as NT. Therefore, it was decided to create a model using these variables and whose output indicated whether or not the volume studied should preserve the segmentation made of the enhancing tumor core. This model was based on a multiple linear regression.

\subsubsection*{Regression models}

Data from the 175 training volumes were used to create the regression models. A first regression model $M_1$ was created using as variables only the relationships $P^j$ shown in eq. \ref{rel} considering the 15 thresholds from 0.15 to 0.85 with steps of 0.05. These variables were called $x_{i}^j$, with $i =$ 1, 2,... 175, indicating the training volume, and $j =$ 1, 2,... 15, corresponding to the threshold used. If for a volume the amount $N_{na}^j$ using some threshold $u_j$ was equal to zero, then the relationship $P^j$ was equal to 10 as shown in eqs. \ref{rel_01}. When a volume really had the enhancing tumor core, then in the regression its dependent variable $y_i$ was chosen equal to 1, while if it really did not have the enhancing tumor core it was chosen equal to 0. Once the regression and the model $ M_1$ was obtained, this was applied to both the training and validation volumes. The regression output was called $y_i'$. A threshold $u_{r1}$ was then chosen such that if $y_i' \geq u_{r1}$, then the volume preserved the segmentation of the enhancing tumor core (assuming this was done), and if $y_i' < u_{r1}$, then the segmentation was removed. Removing the segmentation meant changing the NA label of the voxels to the NCR/TI label. The threshold was chosen so that the best results were obtained with the validation volumes.
Once the above was finished, a second model $M_2$ was created following a procedure similar to the first. The $M_2$ model was applied to the volumes obtained after applying the $M_1$ model. Unlike the $M_1$ model, the variables used in $M_2$ were all the relationships $P^j$, $Q^j$ and $R^j$ shown in eqs. \ref{rel} using the 15 thresholds. These variables were called $x_{i}^j$, $w_{i}^j$ and $z_{i}^j$ respectively, with $i =$ 1, 2,... 175, indicating the training volume, and $j =$ 1, 2,... 15, corresponding to the threshold used. As before, Eqs. \ref{rel_01} were used for a volume when $N_{na}^j$ was equal to 0 using some threshold $u_j$. Then the model $M_2$ was applied to the training and validation volumes. For some volumes the relationships used were Eqs. \ref{rel_01}, because its segmentations were previously removed when the $M_1$ model was applied. As before, a threshold $u_{r2}$ was used with respect to which it was decided to preserve or remove the segmentation of the enhancing tumor core. Volumes that had their segmentation removed after applying the regression models were excluded from the next post-processing technique called \emph{threshold customization with regression}.

\subsubsection{Threshold customization with regression}

For this post-processing technique, a regression model similar to the one used in the previous section was developed. The model had as independent variables the relations of eqs. \ref{rel_0} described above considering the thresholds from 0.15 to 0.95 with steps of 0.05. Different regression models were created using different combinations between the thresholds and the mentioned variables as variables. All possible combinations were made between the thresholds of the form $a$:$b$:$c$, with the lower bound $a =$ 1, 2, ..., 17, the upper bound $b =$ 1 , 2,..., 16 and with steps $c =$ 2, 3,..., 17. Thus, for example, the combination $2$:$3$:$11$ studied all the variables in the same model considering the thresholds numbered 2, 5, 8 and 11. Seven sets of variables shown in Table \ref{rel_comb} were studied separately.

\begin{table}[ht]
	\centering
	
	\begin{tabular}{c|c}
		\hline
		1 & $P_i$ \\
		2 & $Q_i$\\
		3 & $R_i$\\
		4 & $P_i$, $Q_i$\\
		5 & $P_i$, $R_i$\\
		6 & $Q_i$, $R_i$\\
		7 & $P_i$, $Q_i$, $R_i$\\
		\hline
	\end{tabular}
	\caption[Variable sets]{{\bf Variable sets.} The seven sets that were studied separately in the regression models for threshold customization are shown. }
	\label{rel_comb}
\end{table}

As described in the previous section, each variable was measured using a threshold indicated by the superscript $j$. Therefore, for simplification, the same thresholds were used for each of the variables studied. For example, assuming that the relationships $P^j$ and $Q^j$ (set number 4 of Table \ref{rel_comb}) were studied, using the thresholds 1:1:4, then the variables of the regression model were: $\{P^1, P^2, P^3, P^4, Q^1, Q^2, Q^3, Q^4 \}$.
Regarding the dependent variable $y_i$, with $i =$ 1, 2,..., 175 (excluding those whose segmentation was previously removed), this was chosen equal to the ideal threshold with which the i-th volume would have obtained the best possible segmentation of the enhancing tumor core. Of all the models created from combinations between thresholds and variables, the one that achieved the greatest improvement with the validation volumes was chosen. This improvement was compared to the best result obtained by using the same threshold for all volumes. The last post-processing technique used in stage 3 was \emph{refinement of NA}.

\subsubsection{Refinement of NA}

\begin{figure}[ht]
	
	\begin{center} 
		\includegraphics[scale=0.25]{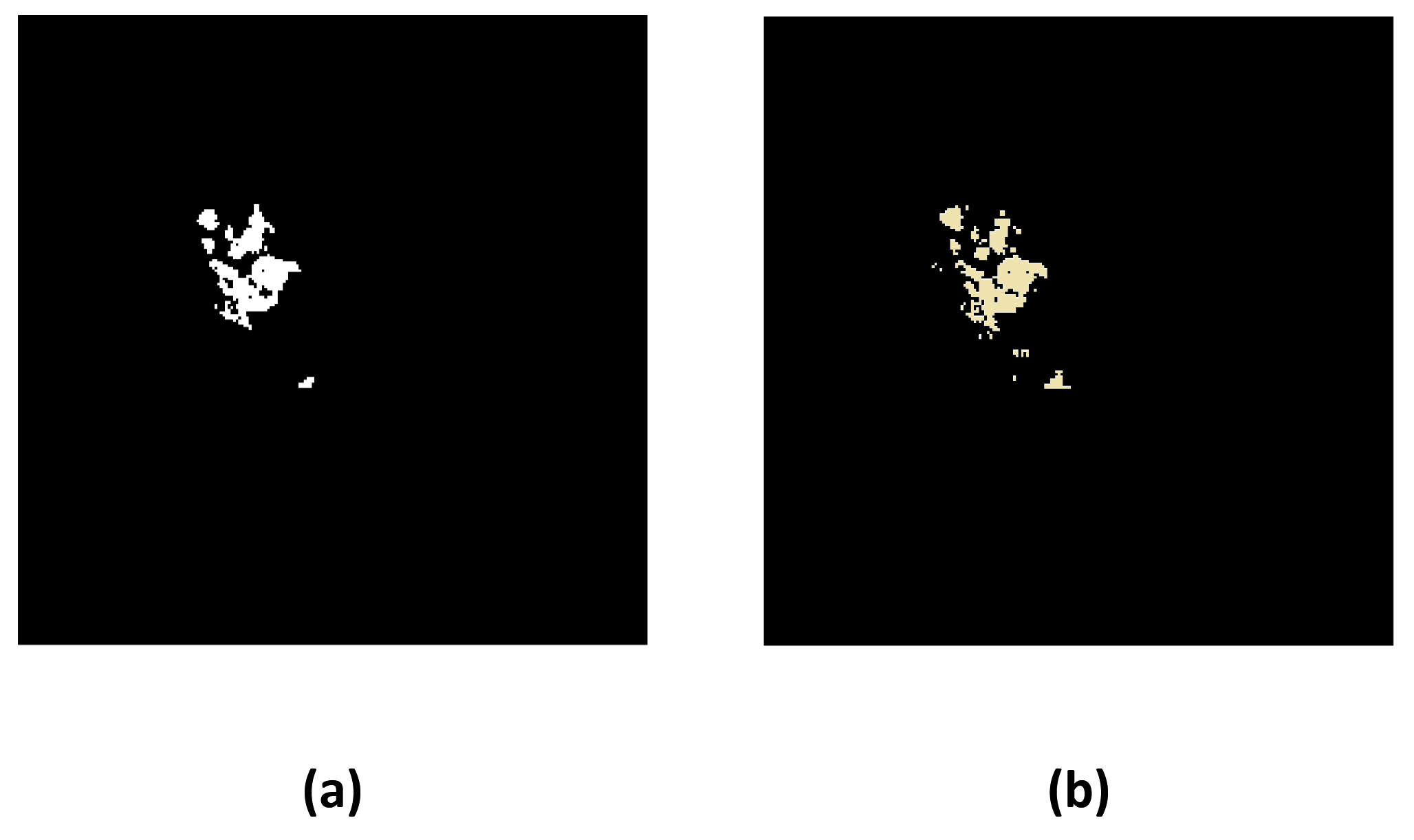} 
		
	\end{center} 
	\caption[Segmentation of the enhancing tumor core.]{{\bf Segmentation of the enhancing tumor core.} (a) After applying the processes described in step 3, an example of the segmentation made of the enhancing tumor core is shown, which was called $S_{ ta}$. (b) Actual segmentation.} 
	\label{et3_final} 
\end{figure}

\begin{figure}[ht] 
	
	\begin{center} 
		\includegraphics[scale=0.35]{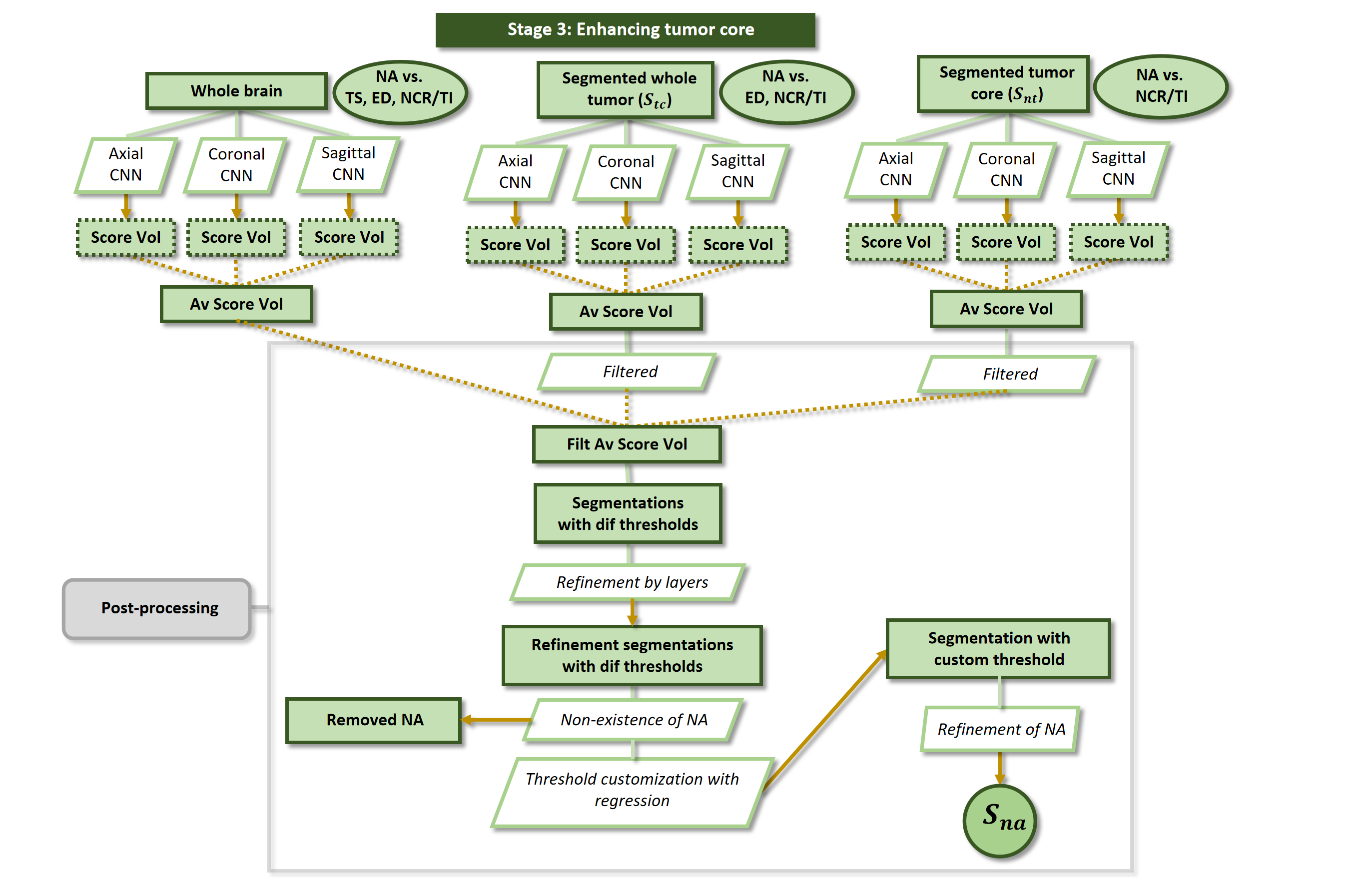} 
		
	\end{center} 

	\caption[Stage 3 summary]{{\bf Stage 3 Summary}. The goal of stage 3 was to segment the "enhancing tumor core". Three different regions were studied: whole brain, segmented whole tumor in stage 1 and segmented tumor core in stage 2. For each region three CNNs were created, each one studying a spatial plane (axial, coronal and sagittal). From the three CNNs an averaged score volume was obtained which was then \emph{filtered}. From the three volumes of the three planes, a single filtered score volume was created. Using different thresholds, segmentations of the enhancing tumor core were obtained to which post-processing \emph{refinement by layers} was applied. Two regression models were created for the process \emph{non-existence of NA}, to decide whether to preserve or remove the segmentation made. The method \emph{customization of the threshold with regression} was applied to each volume. At the end, a \emph{refinement of NA} was applied to obtain the segmentation volumes $S_{ ta} $ del enhancing tumor core.}
	\label{resumen_3} 
\end{figure}

The post-processing \emph{refinement of NA} was pretty much the same as for stage 1 (\emph{refinement of TC}). The only difference was the use of the T\textsubscript{1Gd} modality. All the variables involved in this process were chosen so that the best results were obtained with the validation volumes. The volume of the segmentation of the enhancing tumor core was called $S_{na}$ (Fig. \ref{et3_final}). As in the previous stages, the segmentations $S_{na}$ and the real segmentations were compared through the calculation of the Dice coefficient. Finally, Fig. \ref{resumen_3} shows the summary of stage 3.

\subsection{Final segmentations}

The slices $S_{tc}$ and $S_{nt}$, built in stages 1 and 2 respectively, were not the final slices. In stage 2, due to various post-processing techniques, some new voxels included in the tumor core segmentation $S_{nt}$, were initially not part of $S_{tc}$. Since $S_{nt}$ should have been contained in $S_{tc}$, then the new voxels contained in $S_{nt}$ were added to $S_{tc}$. Similarly, in stage 3 due to various post-processing techniques, some new voxels included in the segmentation of the enhancing tumor core $S_{na}$ were not part of $S_{nt}$ nor of $S_{tc} $. Then these new voxels were also included in $S_{nt}$ and $S_{tc}$. Therefore, including all the missing voxels, the final segmentations of the whole tumor $S_{tc}'$ and the tumor core $S_{nt}'$ were created. The segmentation of the enhancing tumor core $S_{na}$ did not undergo changes, so this was its respective final segmentation.

\section{Results}

\subsection{Stage 1}

\begin{figure}[ht] 
	
	\begin{center} 
		\includegraphics[scale=0.3]{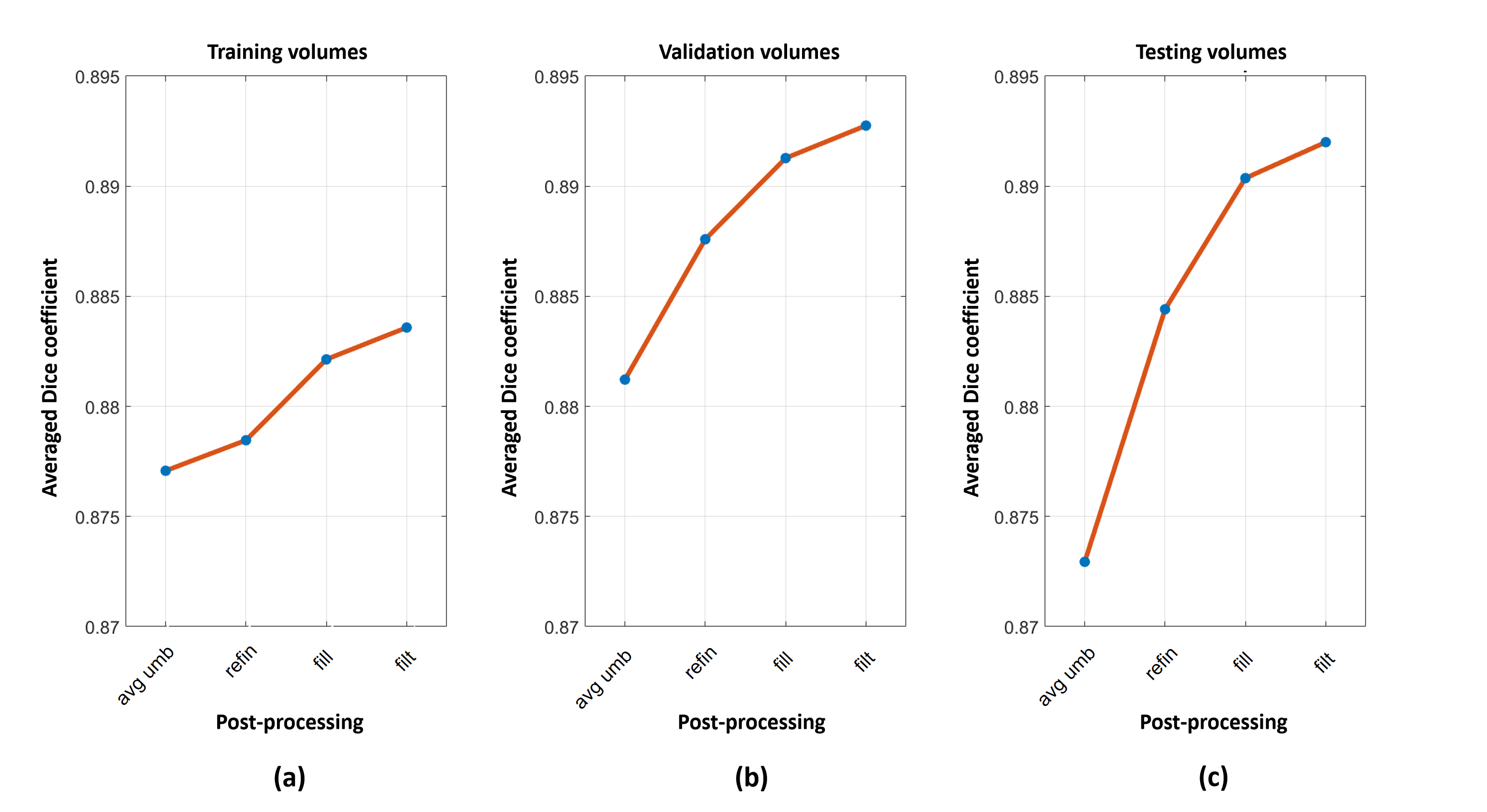} 
		
	\end{center} 
	\caption[Results of all processes of stage 1]{{\bf Results of all processes of stage 1.} The average Dice coefficients achieved without post-processing, using the same threshold equal to 0.65 (avg umb), are shown . The results obtained after applying the processes of \emph{refinement of TC} (refin), \emph{fill by addresses} (fill) and \emph{filtered} (filt) are indicated, for training (a), validation (b) and testing (c) data. } 
	\label{seg_all_2} 
\end{figure}

The best result for the validation volumes without post-processing was obtained using the threshold equal to 0.65, reaching the average Dice coefficient equal to 0.8812. Also the training and testing volumes obtained the best results with the same threshold, being equal to 0.8771 and 0.8729 respectively.
For the \emph{refinement of TC} technique, the threshold number $U_p = $ 11 was chosen (since $u_p =$ 0.65). In addition, the following parameters were chosen: $U_{sup} = $6, $U_{up} = $5 and $U_{donw} =$3. Regarding the percentages, these were: $P_{down} = $20 and $P_{up} = $30. These values were chosen such that the greatest improvement was achieved with the validation volumes. With the above, the average Dice coefficients for the training, validation and testing volumes were equal to 0.8785, 0.8876 and 0.8844 respectively.
Subsequently, using the \emph{address fill} technique the results for the three data subsets were equal to 0.8821, 0.8913 and 0.8904 respectively. Finally, applying the \emph{filtered} technique, the results obtained were equal to 0.8836, 0.8928 and 0.8920, respectively. In Fig. \ref{seg_all_2} all the previous results were grouped.

\subsection{Stage 2}

\begin{figure}[ht] 
	
	\begin{center} 
		\includegraphics[scale = 0.3]{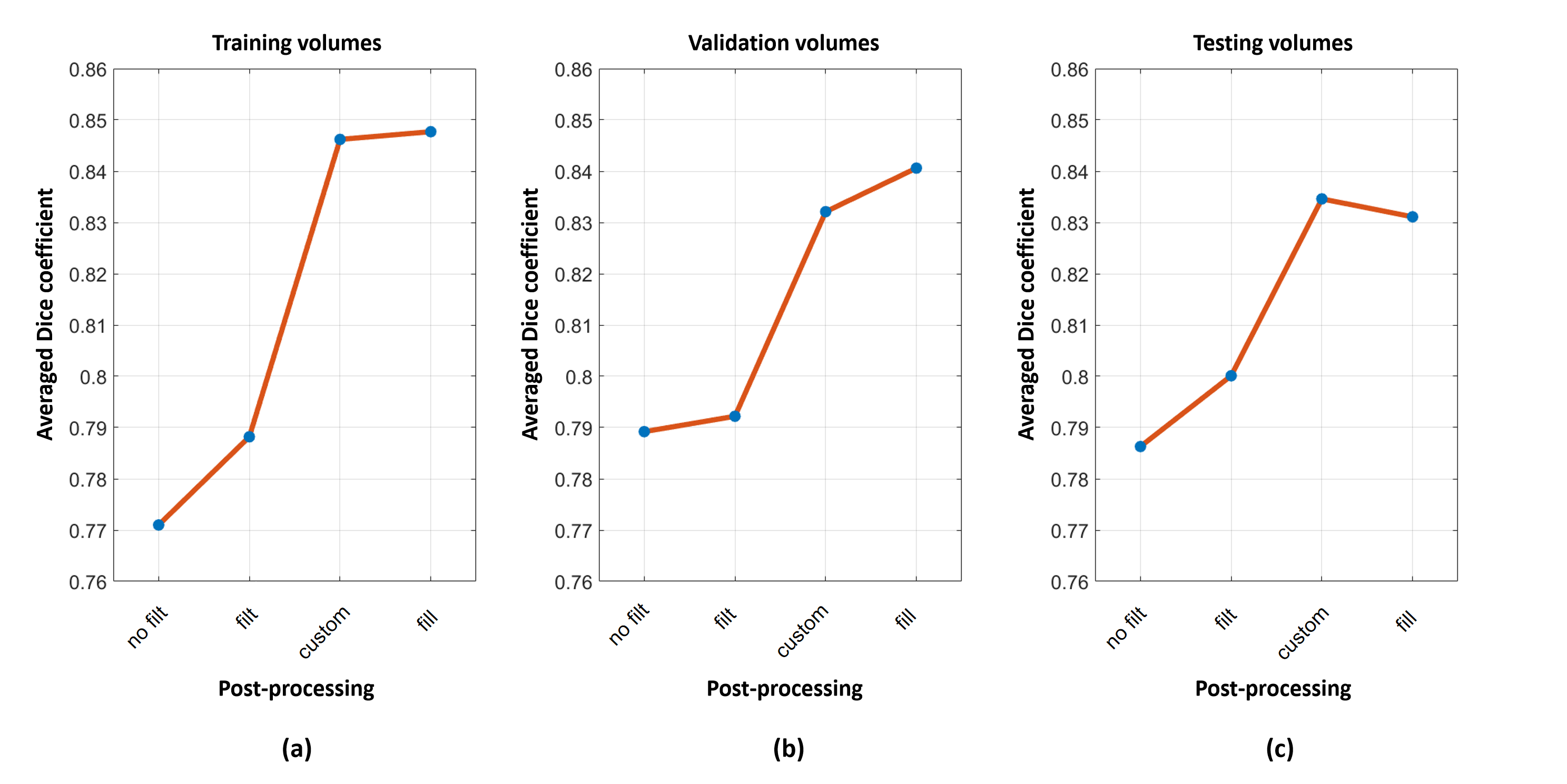} 
		
	\end{center} 
	\caption[Results of all the processes of stage 2]{{\bf Results of all the processes of stage 2.} The average Dice coefficient of the segmentations is indicated when using the threshold equal to 0.6 without filtering the score volumes (no filt) and using the threshold equal to 0.5 after applying the \emph{filtered} technique (filt). Those achieved after applying the \emph{threshold customization with RNN} (custom) technique are shown. Finally, the results are shown after applying the technique \emph{fill by planes} (fill). All of the above for the training (a), validation (b) and testing (c) volumes.  } 
	\label{res_proc_1_4_2_all} 
\end{figure}

Before applying the filter, the threshold at which the validation volumes obtained the highest average Dice coefficient was equal to 0.6. Using this threshold the results for the training, validation and test volumes were equal to 0.7710, 0.7892 and 0.7863 respectively. In the \emph{filtered} post-processing technique, the median filter was applied to the score volumes a total of 20 times. Different thresholds were then used to obtain the respective segmentations. For the validation volumes, the best result was obtained using the threshold equal to 0.5. Using this same threshold, the training, validation, and testing volumes reached average Dice coefficients equal to 0.7882, 0.7922, and 0.8001, respectively. These results are improved by applying the \emph{threshold customization with RNN} technique. Using the custom thresholds, for the validation volumes the average Dice coefficient was equal to 0.8321. For the training and testing volumes this was equal to 0.8462 and 0.8346 respectively. Finally, after applying the \emph{fill by planes} technique, the final average Dice coefficients for the training, validation, and testing volumes were equal to 0.8477, 0.8406, and 0.8311, respectively. Fig. \ref{res_proc_1_4_2_all} groups the described results.

\subsection{Stage 3}

Taking the results of the validation volumes as a reference, when no filter was applied and no refinement was performed, the best result was obtained using the threshold equal to 0.6, obtaining an average Dice coefficient equal to 0.7415. After applying the \emph{Gaussian filter} technique, the best result was obtained with the threshold equal to 0.5, increasing the coefficient to 0.7455. For the \emph{refinement by layers} technique, two internal layers and two external layers were chosen with respect to the threshold layer used, in addition to a cube of dimensions 3$\times$3$\times$3. Applying this technique and using a threshold equal to 0.6, the average Dice coefficient was equal to 0.7476. Applying all of the above, the training volumes had the average Dice coefficients equal to 0.7029, 0.7151 and 0.7219, respectively. The test volumes obtained the coefficients equal to 0.7610, 0.7779 and 0.7798 respectively. It should be remembered that these results included volumes that obtained Dice coefficients equal to zero, since they did not really have the enhancing tumor core and erroneously had a segmentation of this. Table \ref{M_1} shows the coefficients associated with the variables used by the model $M_1$ and Table \ref{M_2} shows the coefficients of the variables used by the model $M_2$.

\begin{table}[ht]
	\centering
	
	\begin{tabular}{cc|cc}
		\hline 
		\textbf{Variable}  & \textbf{Coefficient} &   \textbf{Variable}  & \textbf{Coefficient}   \\				
		\hline  
		$P^1$ & -0.1339 &$P^9$ & -0.0363\\
		$P^2$ & 0.1634  &$P^{10}$ & -0.0136\\
		$P^3$ & 0.0029  &$P^{11}$ & 0.0241 \\
		$P^4$ & -0.0162&$P^{12}$ & -0.0057 \\
		$P^5$ & -0.0129&$P^{13}$ & 0.00009 \\
		$P^6$ & 0.0009&$P^{14}$ & -0.0004\\
		$P^7$ & 0.0036&$P^{15}$ & -0.0004\\
		$P^8$ & 0.0353&$cte$ & 0.9313\\
		\hline
	\end{tabular}
	\caption[Variables and coefficients of the regression model \emph{M}$_1$]{{\bf Variables and coefficients of the regression model \emph{M}$_1$.}}
	\label{M_1}
\end{table}

\begin{table}[ht]
	\centering
	
	\begin{tabular}{cc|cc|cc}
		\hline 
		\textbf{Variable}  & \textbf{Coefficient} &   \textbf{Variable}  & \textbf{Coefficient}   \\				
		\hline
		$P^1$&0.5326 &
		$P^2$&-5.5115 &
		$P^3$&7.5868\\
		$P^4$&-2.4737&
		$P^5$&88.4441&
		$P^6$&-178.3476\\
		$P^7$&20.7119&
		$P^8$&116.0197&
		$P^9$&1.4187\\
		$P^{10}$&-121.8615&
		$P^{11}$&11.5613&
		$P^{12}$&-8.5484\\
		$P^{13}$&265.4135&
		$P^{14}$&-225.1223&
		$P^{15}$&31.3617\\
		$Q^1$&0.0088&
		$Q^2$&-0.6474&
		$Q^3$&0.7026\\
		$Q^4$&-0.1670&
		$Q^5$&0.0250&
		$Q^6$&0.0509\\
		$Q^7$&-0.2294&
		$Q^8$&0.0816&
		$Q^9$&0.1610\\
		$Q^{10}$&-0.2286&
		$Q^{11}$&0.1961&
		$Q^{12}$&-0.08586\\
		$Q^{13}$&0.0136&
		$Q^{14}$&-0.0016&
		$Q^{15}$&-0.0004\\
		$R^1$&0.0781&
		$R^2$&-0.0927&
		$R^3$&-0.0674\\
		$R^4$&0.0067&
		$R^51$&-47.7644&
		$R^6$&106.6641\\
		$R^7$&-28.7388&
		$R^8$&-77.8702&
		$R^9$&31.4158\\
		$R^{10}$&67.0931&
		$R^{11}$&-27.6441&
		$R^{12}$&27.7117\\
		$R^{13}$&-167.4537&
		$R^{14}$&130.8674&
		$R^{15}$&-14.7525\\
		$cte$&1.2064& & & & \\
		\hline
	\end{tabular}
	\caption[Variables and coefficients of the regression model \emph{M}$_2$]{{\bf Variables and coefficients of the regression model \emph{M}$_2$.}}
	\label{M_2}
\end{table}

For the model $M_1$, the threshold $u_{r1} = 0.5$ was used, and similarly for the model $M_2$ the threshold $u_{r2} = 0.5$ was used. Applying only the $M_1$ model, 8 out of 175 training volumes (5\% of the total) had an erroneous prediction. Specifically, these 8 volumes did not really have an enhancing tumor core, and the result of the $M_1$ model indicated that segmentation of this subregion should be preserved if it had been done. However, when model $M_2$ was subsequently applied to the results of model $M_1$, the erroneous predictions were corrected, reaching 100\% accuracy.
Similarly, for the validation and testing volumes, the erroneous predictions made by the $M_1$ model were corrected by the subsequent application of the $M_2$ model, obtaining 100\% accuracy in both sets. Therefore, when a volume did not have the enhancing tumor core, then the segmentation made of this subregion was removed (if it existed), obtaining for this case a Dice coefficient equal to 1.
After applying the models $M_1$ and $M_2$ and using a threshold equal to 0.6 (with which the validation volumes reached the best results), average Dice coefficients equal to 0.8134, 0.7988 and 0.8054 were obtained for the training, validation and testing volumes respectively.
Regarding the model created for the post-processing technique \emph{threshold customization with regression}, only the variables $P^j$, $Q^j$ and $R^j$ were studied with $j = $ 1, 3 and 5 (corresponding to the thresholds equal to 0.15, 0.25 and 0.35 respectively). With this combination the validation volumes achieved the best results. Table \ref{coef_person_1_4} shows the coefficients of the model. After applying the \emph{threshold customization with regression} technique, the average Dice coefficients for the training, validation, and testing volumes were equal to 0.8162, 0.8050, and 0.8078, respectively.

\begin{table}[ht]
	\centering
	
	\begin{tabular}{cc|cc}
		\hline 
\textbf{Variable}  & \textbf{Coefficient} &   \textbf{Variable}  & \textbf{Coefficient}   \\				
		\hline
$P^1$ & 0.7334 & $Q^5$ &-0.0048  \\
$P^3$ & -0.6043 & $R^1$ &-0.0125 \\
$P^5$ & -2.2125 & $R^3$ &0.0348 \\
$Q^1$ & -0.0258 & $R^5$ &1.3315 \\
$Q^3$ &0.0114 & $cte$ &0.5418 \\
		\hline
\end{tabular}
\caption[Coefficients of the model used in the technique \emph{threshold customization with regression}]{{\bf Coefficients of the model used in the technique \emph{threshold customization with regression.}}}
\label{coef_person_1_4}
\end{table}

\begin{figure}[ht] 
	
	\begin{center} 
		\includegraphics[scale = 0.3]{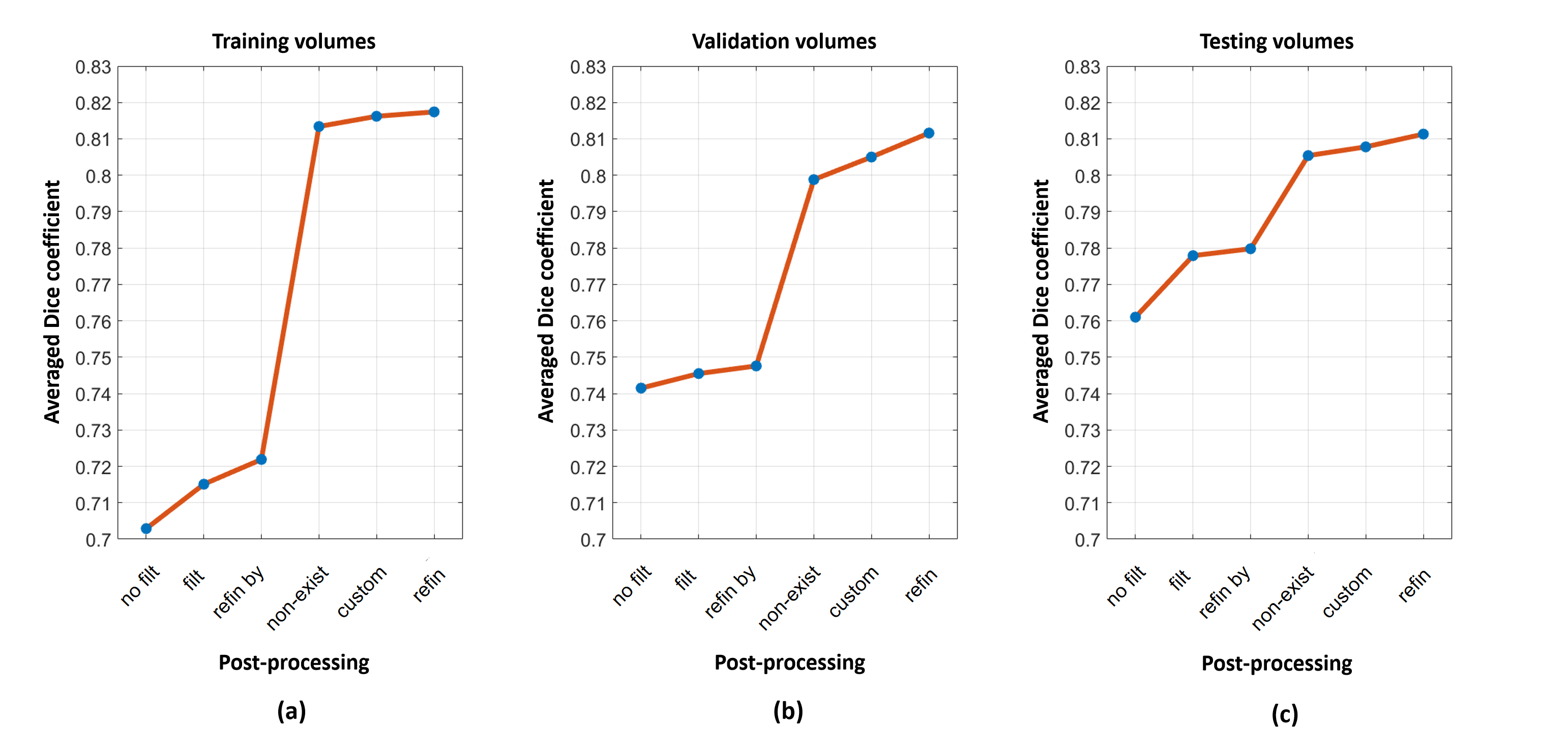} 
		
	\end{center} 
	\caption[Results of all the processes of stage 3]{{\bf Results of all the processes of stage 3.} The average Dice coefficient is indicated after applying the following processes. Using a threshold equal to 0.6 without filtering (no filt). Occupying a threshold of 0.5 after applying the \emph{filtered} technique (filt). Considering a threshold equal to 0.6 and applying the \emph{refinement by layers} technique (refin). By applying the process \emph{non-existence of NA} (non-exist). After considering the \emph{threshold customization with regression} technique (custom). And finally when the last process of \emph{refinement of NA} (refin) was performed. All of the above for the training (a), validation (b) and testing (c) volumes.} 
	\label{person_umb_1_4_all} 
\end{figure}

For the \emph{refinement of NA} technique, each volume used the reference threshold $U_p$ assigned by applying the \emph{threshold customization with regression technique.} Then, for each volume, the following thresholds were considered: $U_ {sup} = U_{p} + 3$, $U_{up} = U_p + 1$ and $U_{down} = U_p - 1$. On the other hand, the percentages were chosen: $P_{down} = $ 35 and $P_{up}$ = 60. As on other occasions, these values were chosen in such a way that the validation volumes achieved the greatest improvement in their segmentations. So, after performing the \emph{refinement of NA}, the average Dice coefficients for the training, validation, and testing sets were equal to 0.8174, 0.8116, and 0.8113, respectively. Fig. \ref{person_umb_1_4_all} groups all the described results of stage 3.

\subsection{Final segmentations}

After adding the missing voxels in the respective segmentations due to post-processing techniques, the results of the final segmentations for the training, validation and testing volumes, considering the three regions of the tumor called whole tumor, tumor core and enhancing tumor core, are indicated in Table \ref{seg_final}. For comparison, Table \ref{comp} shows the final results of the segmentations made for the testing volumes and what was obtained by the first \cite{1}, second \cite{2} and third \cite{ 3_2019} place in the BraTS challenge in its 2019 version \cite{3}.

\begin{table}[ht]
	\centering
	
	\begin{tabular}{|c|c|c|c|c|c|}
		\hline 
		\textbf{Sets} & \textbf{TC (stage 1)} & \textbf{TC (final)}& \textbf{NT (stage 2)} & \textbf{NT (final)}& \textbf{NA (stage 3 and final)}  \\
		\hline
		Training & 0.8836 & 0.8842 & 0.8477 & 0.8506 & 0.8174 \\
		Validation & 0.8928 & 0.8938 & 0.8406 &  0.8495 & 0.8116 \\
			Testing & 0.8920 & 0.8934 & 0.8311 & 0.8376 & 0.8113  \\	
	
		\hline
	\end{tabular}
	\caption[Final results for segmentations]{{\bf Final results for segmentations.} The average Dice coefficients of the training, validation and testing volumes are shown. These were those obtained for the whole tumor (TC), tumor core (NT) and enhancing tumor core (NA) regions. Both the results achieved at the conclusion of stage 1, stage 2 and stage 3 are indicated, as well as the final results when adding the missing voxels due to the post-processing techniques (final). }
	\label{seg_final}
\end{table}

\begin{table}[ht]
	\centering
	
	\begin{tabular}{|c|c|c|c|}
		\hline 
		\textbf{} & \textbf{TC} &  \textbf{NT} & \textbf{NA}  \\
		\hline
		This work & 0.8934 & 0.8376 & 0.8113  \\
		First place & 0.8880 & 0.8370 & 0.8327  \\
		Second place & 0.8830 & 0.8610 & 0.8100 \\
		Third place & 0.89 & 0.83 & 0.81 \\
		\hline
	\end{tabular}
	\caption[Comparison]{{\bf Comparison.} Results obtained with the testing volumes in this work, and those obtained by the participants of the BraTS 2019 challenge who reached first, second and third place.}
	\label{comp}
\end{table}

\section{Discussion}

Different CNNs were trained for glioma segmentation, with simple architectures and studying a relatively small number of samples obtained from conventional MRI. To improve the results of the created CNNs, original post-processing techniques were proposed, thus reaching the state of the art in the automatic segmentation of the histological subregions of gliomas. These subregions were the whole tumor, the tumor core, and the enhancing tumor core.

\subsection{Justification of post-processing techniques}

The currently most widely used image analysis technique for glioma segmentation is based on CNNs. An important aspect for its implementation is the high hardware requirement, mainly due to the large amount of data used for its training, as well as the execution of complex algorithms for its construction. In the present work, a conventional computing system (Intel Core i7-4790, 32GB of RAM, GeForce RTX 2060, 6GB VRAM, Windows 7) was used for the development of the CNNs described above. Therefore, there were certain limitations that had an impact on the way the CNNs were built and trained, as well as the results they achieved. For example, the networks were trained with a relatively small amount of input data. These data were patches of small size in two dimensions. Also the networks were formed by a reduced number of layers and filters. Therefore, the segmentations made, when only the proposed CNNs were used, did not achieve sufficiently outstanding results compared to what is reported in the literature. Considering the above, there was a need to improve the results obtained by the networks. Thus, the main contribution of the presented work was the invention, adaptation and application of different original post-processing techniques that improved to a lesser or greater extent the segmentations made by the CNNs. Among the proposed techniques were the following: \emph{center}, \emph{connected neighbors}, \emph{refinement of TC and NA}, \emph{refinement by layers}, \emph{fill by directions}, \emph {fill by planes}, \emph{filtered (median or Gaussian)}, \emph{non-existence of NA} (models $M_1$ and $M_2$) and \emph{threshold customization with RNN and regression) }. As could be observed in Figs. \ref{seg_all_2}, \ref{res_proc_1_4_2_all} and \ref{person_umb_1_4_all}, the improvement of the results after applying the post-processing techniques was evident. Thus, the implementation of these techniques was justified. In the following sections, the contributions of each technique will be described, in addition to the assumptions from which they were based.

\subsubsection{Center}

This technique had the hypothesis that the highest concentration of voxels connected to each other and with scores greater than 0.95 correctly belonged to the whole tumor. At least with the data studied in the present work and considering the aforementioned hypothesis, the \emph{center} technique always correctly located the approximate geometric center of all the considered whole tumors. Unlike other semi-automatic methodologies in which the user must manually provide the position of the tumor \cite{semi_auto_located}, in the present work through the \emph{center} technique, this task was successfully done automatically.
 
 \subsubsection{Connected neighbors}

Different segmentations were performed by choosing a threshold or interval of thresholds, so that if the voxel score was within this, then it was included in the segmentation. However, there were voxels whose score did meet the threshold but were not really part of the whole tumor. These voxels were usually found far away and isolated (i.e., disconnected) from the highest concentration of voxels that did correctly belong to the tumor. Therefore, starting from the chosen voxel through the \emph{center} technique, the segmentations (using some threshold) included only the voxels that were connected to each other. Since the use of different thresholds was explored during the work, the task of locating all the voxels connected to each other became a computationally inefficient process. This is due to the number of voxels to consider even repetitively. In the post-processing technique called \emph{connected neighbors}, different successive segmentations were made, each one formed by voxels that met the same threshold condition and that were also connected to each other. In order to make the connection between previous and later segmentation voxels, only the voxels that were connected to each other and connected to 50\% of the last voxels added to a previous segmentation were searched. All voxels located on the "surface" of a previous segmentation were assumed to be within 50\% of the last added voxels. This assumption was not necessarily true, due to the shape and structure of the tumor. So it could have caused some voxels of interest to have been erroneously excluded from the segmentation. However, despite this, the objective of making the segmentation process more efficient was achieved. In addition, better results were obtained compared to not having applied the technique.

\subsubsection{Refinement of TC and NA}

In this technique it was assumed that the highest frequency intensities measured in a particular MRI modality, considering only the voxels that had a high score, were more representative of the tumor region to be segmented. Remembering that a high score indicated a higher probability that a voxel belonged to the tumor, then this was the justification for choosing these voxels. Reviewing the protocol for manual segmentation of gliomas described by the organizers of the BraTS \cite{brats2} challenge, the FLAIR modality was the modality from which the segmentation of the whole tumor began. In the opinion of the organizers and based on the experience of the experts involved, the FLAIR and T\textsubscript{1Gd} modalities were the most useful for performing these segmentations. Therefore, these were the reasons for choosing the FLAIR modality in the refinement for stage 1. In stage 3, since the region to segment was the enhancing tumor core, the T\textsubscript{1Gd} modality was a choice natural. On the other hand, it should be mentioned that the choice of these modalities also had empirical support. That is, applying the \emph{refinement} technique in these modalities the greatest improvements in the segmentations were obtained.

\subsubsection{Refinement by layers}

It was assumed that the score volume could be divided into layers, each containing voxels whose scores were within a certain threshold. In this technique only new voxels were added to the segmentation made. This was because the enhancing tumor core usually had relatively small dimensions, making it more convenient to refine by adding voxels instead of excluding them. On the other hand, the added voxels that belonged to the outer layers of the segmentation were restricted to being located at most a certain distance from the voxels of the inner layers (these being the ones that were part of the initial segmentation). This to only add voxels close to the initial segmentation. Although segmentations cannot be divided into layers because tumors do not have a uniform shape, the description of \emph{refinement by layers} is only conceptual to help understand its methodology, being equally useful and applicable even if tumors have diverse structures.

\subsubsection{Fill by directions and planes}

The technique called \emph{fill by directions} was used in stage 1. It was a simple and convenient process for solving the problem of ``hollows'' within segmentations. It should be mentioned that regions that had formed a ``hollows'' were not always filled. Due to the linear nature of the scans, spaces were sometimes filled by forming lines within the segmentation and spaces with the same shape were also left unfilled. Despite this, the objective of improving the segmentation results was achieved, and with the subsequent application of the \emph{filtered} technique, the problem of filling forming lines was usually solved. The technique called \emph{fill by planes} was used in stage 2. This technique was similar to the previous one, with the main difference that the process was stricter since it had a greater number of iterations. The reason for this was that the tumor core had smaller dimensions than the whole tumor and it was convenient to use the \emph{fill by planes} technique, in order to perform the filling more strictly.

\subsubsection{Filtered}

The \emph{filtered} technique applied the well-known median and Gaussian filters. These filters generated a smoothing in the segmentations made and were applied for different reasons in the stages of the work. In stage 1, the median filter was applied to the segmented volume. This helped to fill in the missing hollows within the regions of interest. This was achieved since during smoothing the voxels in the hallows acquired the properties of the surrounding voxels, which implied their filling. In stage 2, the median filter was applied to the score volumes to exclude small groups of distant voxels that should not have been part of the segmentation. In stage 3, the Gaussian filter was also applied to the score volume and to exclude erroneous voxels. This was possible since the smoothing filters reduced the scores of the small groups of voxels mentioned, so that using an appropriate threshold, their exclusion in the segmentation was facilitated. The reason for using the Gaussian filter was that in stage 3 the most difficult region, the enhancing tumor core, was segmented. This then required a more elaborate filter. Depending on the size of the Gaussian filter, it gave the voxels a different weight (based on a Gaussian distribution) in the final result. As described during the methodology, for the volume scores $V_{nt}$ and $V_{tc}$ a value of $\sigma$ equal to 0.5 and 1.5 was used, respectively. The sizes of these filters were equal to 3$\times$3$\times$3 and 7$\times$7$\times$7 voxels, respectively. For the volume $V_{cc}$ no filter was applied.

To explain the reason for choosing the size of the filters, the following must first be remembered. When the study region of a CNN was restricted to the segmented region in a previous stage, only the labels that in principle should have been contained in it were considered. For example, CNN 1 studied the segmented region of the tumor core and considered only the NA and NCR/TI labels. However, since no segmentation was perfect, then in that region there could be voxels with labels that were not considered in the network. Thus, in the previous example, there could have been voxels whose real labels were ED or TS. Something similar happened with CNN 2, which studied the segmented region of the whole tumor, considering only the ED, NA and NCR/TI labels, when voxels with the real TS label could have erroneously existed. Regarding CNN 3, there was no such problem, since its study region was the entire volume and it considered all possible labels. On the other hand, and as explained before, the usefulness of the smoothing filters was to reduce the value of the scores of small sets of voxels that were isolated and far away. Considering all of the above, the filter size for the volume $V_{tc}$ was larger than the filter for the volume $V_{nt}$. The reason was that the volume $V_{tc}$ studied a larger region compared to the volume $V_{nt}$, so $V_{tc}$ could contain more voxels with high scores with a real label different from NA. Regarding the volume $V_{cc}$, there were two reasons that justified not performing any filter on it. The first was that volume $V_{cc}$ could contain more precise information compared to volumes $V_{tc}$ and $V_{na}$, since CNN 3 considered all possible labels. The second was that, even though in volume $V_{cc}$ there could be more sets of voxels far away and isolated from the region of interest and with erroneous labels, when performing the average with the other two volumes, these were excluded. This happened because if a remote voxel in $V_{cc}$ erroneously had a high score, in $V_{tc}$ and $V_{na}$ it had scores equal to 0 since it was not part of their study region. Thus, by averaging its three scores, the final score had a low value, and using an appropriate threshold, it was therefore excluded from the segmentation.

\subsubsection{Non-existence of NA}

A common problem is the segmentation of the enhancing tumor core to volumes that did not really have this region. Several works have reported post-processing to resolve the issue. This commonly consists of excluding the segmented region if the number of voxels that formed it was less than a certain threshold chosen individually for each volume \cite{1_2019,2_2018}. In this work, it was observed that, when a volume did not really have this region, then the number of voxels of the enhancing tumor core was relatively lower than the voxels of the tumor core. Considering this, different relationships (eqs. \ref{rel}, \ref{rel_0} and \ref{rel_01}) were proposed and used as variables in two multiple linear regression models $M_1$ and $M_2$. These were applied to the volumes to determine whether or not they preserved the enhancing tumor core. After applying the models, an accuracy of 100\% was reached. This meant that, after applying the models to the training, validation and testing volumes, all volumes that did not actually have the enhancing tumor core were correctly identified and the region was removed. And on the other hand, all the volumes that really did have the region were identified and their segmentation was preserved. Observing the results shown in Fig. \ref{person_umb_1_4_all}, the application of the \emph{inexistence of NA} technique was the one that generated the greatest improvement in the segmentations.

\subsubsection{Threshold customization with RNN and regression}

For a particular volume, it was observed that there was a threshold such that the best possible segmentation could be obtained, being also different compared to other volumes. Therefore, it was decided to develop tools to help customize the choice of threshold. For this, in stage 2 a recurrent neural network (RNN) was trained, while in stage 3 a linear regression model was created. Recurrent neural networks (RNNs) are designed for the study of ordered input sequences. In stage 2, the variables used to personalize the threshold were the most frequent high scores, obtained from the histogram of the score volumes. These variables were sequences of ordered values, therefore the use of an RNN for threshold customization was convenient and justified. On the other hand, being a recurrent network, it is known that the number of parameters to be adjusted is less compared to a conventional neural network, so that the performance of the algorithms is improved. For stage 3, due to the success obtained with the \emph{non-existence of NA} technique which studied the relationships of eqs. \ref{rel}, it was decided to use these same relations to create another linear regression model for threshold customization. It should be mentioned that the number of training samples for the RNN was limited to the number of training volumes, that is, 175. Evidently this was an insufficient amount to train an RNN. However, observing Fig. \ref{res_proc_1_4_2_all}, even when the network was trained with a reduced number of samples, the improvement obtained after applying this technique was useful and outstanding. In any case, for the models used in the customization processes, it will be necessary to develop them with a greater amount of data in future works.

\subsubsection{Reported parameter values}

As can be seen, the trained neural networks, the built linear regression models and the proposed post-processing techniques depended on the choice of different parameter values. As explained before, the parameter values were chosen such as to improve the segmentations of the validation volumes. The purpose of the presented work was not to demonstrate that the choice of these values will allow obtaining good results for any other database. In other words, it is recognized that the reported parameter values are mainly related to the validation volumes studied, and probably if another set of data had been considered, then the values could have been different. The goal of the work was to demonstrate that the proposed methodologies can be useful for the glioma segmentation task. As in any other novel methodology, the determination of the parameter values that allow the generalization of the results will occur after other study groups apply the techniques proposed in other databases. This should then be repeated, increasing the number of samples, until consistency is achieved between the results of different groups using the same parameter values. Despite the above, it should be remembered that the choice of the parameter values reported in this work not only allowed obtaining good results with the validation volumes, but also good results were obtained with the training and testing sets. With the three data sets, the same behavior of improvement in the segmentations was practically always observed after applying the post-processing techniques. Therefore, taking as reference Figs. \ref{seg_all_2}, \ref{res_proc_1_4_2_all} and \ref{person_umb_1_4_all}, it can be ensured that, at least with the data studied in this work, there was reproducibility and generalization.

\subsection{Applicability to other problems}

The segmentation work consisted in the training of CNNs and the proposal of different post-processing techniques. Regarding the CNNs, having been trained with the gliomas, then the networks learned the characteristics and patterns of the tumors. So, the possibility of implementing the networks to other segmentation problems could not be immediate. However, the different architectures and part of the adjusted parameters could be used to be trained using the data associated with the new segmentation task (transfer learning). Regarding the post-processing techniques, similarly to the CNNs, the parameters reported in this work are exclusively related to the glioma segmentation problem. On the other hand, the techniques could also be applied in solving other problems. For this, the appropriate parameter values should be obtained to improve the segmentations made by the CNNs trained for the new problem. Although, until now, the reported methodology has not been applied in other segmentation problems, there is no reason that prevents the implementation of these processes even to other MRI modalities different from the conventional ones. Many of the intuitive ideas that led to propose the reported post-processing techniques could be equally valid for segmentation in different types of medical images and regions of interest. This will be a reason for future work on other segmentation tasks.

\subsection{Comparison with the state of the art}

After presenting the discussion about the CNNs and the different post-processing techniques, a comparison is described between the work done and the state of the art in the segmentation of gliomas mainly represented by the winning works of the BraTS challenge of the year 2019.
First, a summary of the work done by the winners will be presented. Later, a general comparison between the main characteristics of this study and those of the winners mentioned, also including the participants of the challenge of the years 2018 and 2019, will be shown.

\subsubsection{First place}

Inspired by the cascade strategy, Jian et al. proposed two stages of a CNN variant known as 3D U-Net \cite{1}. In the first stage, U-Net had the goal of making a rough prediction. Then, in the second stage, they increased the width of the network and used two decoders such that they increased their performance. The second stage was included to refine the prediction map through the concatenation of a preliminary prediction map and the original input to use autocontext. The network inputs were patches of the four available MRI modalities, such that they formed a volume of dimensions 4$\times$128$\times$128$\times$128. This volume was introduced in the network of the first stage. The resulting coarse segmentation map was then concatenated with the original volume and fed into the second stage network. Due to the size of the input patch, they only used a minibatch of size equal to 1. After applying their model to the test data set of the challenge, they obtained average Dice coefficients equal to 0.8880, 0.8370 and 0.8327 for the whole tumor, the tumor core and the enhancing tumor core, respectively.

\subsubsection {Second place}

Zhao et al. combined three sets of what they called ``tricks'' applied to tumor segmentation \cite{2}. The first consisted of data processing methods (data sampling, training with random patch sizes, and semi-supervised learning), method design models (architecture design and result fusion), and process optimization (warm-up). and multitasking learning). The authors built a U-Net architecture that took different patch sizes and number of samples for the batches, which changed randomly at each training iteration. Then, in order to implement semi-supervised learning, they used the manual labels included in the database in the first training iteration. Different ``student'' type models to predict labels were combined. A new training set that combined the manual labels with those predicted by the model was created and the same procedure was repeated in the second iteration and so on.
Another ``trick'' implemented was the introduction of a so-called self-assembly architecture, in which the predictions made at the different scales of the U-Net architecture were joined to generate a final prediction. The patch size was randomly chosen from 64, 80, 96, 112, 128, and 144, with the batch sizes equal to 15, 8, 4, 2, 1, and 1, respectively. Regarding the fusion of results, they performed a cross-validation of 5 iterations and averaged the respective 5 resulting models in order to assemble a final model. Another fusion was performed by combining the different predictions assigned to the same voxel. These predictions were obtained since the same voxel could be part of different patches. Finally, on the optimization methods, they applied a gradual increase in the learning range in each iteration and carried out a reorganization of the classes to be labeled. The combination of all the described tricks allowed them to reach average Dice coefficients equal to 0.910, 0.835 and 0.754 for the whole tumor, the tumor core and the enhancing tumor core, respectively.

\subsubsection {Third place}

McKinley et al. presented a network architecture for semantic segmentation, incorporating dense blocks and dilated convolutions \cite{3_2019}. This architecture is based on one previously developed by the authors, called DeepScan. In this network the authors added a lightweight attention mechanism and replaced batch normalization with example-based normalization. The network was composed of two-dimensional convolutions, but was built from three-dimensional convolutions included in the initial layers. Thus, the receptive field (size of the convolution filters) of the network was anisotropic (it varied depending on the chosen direction). Therefore, the network was trained on the sagittal, coronal, and axial planes. The final result of the network was obtained by assembling the result of the three planes. The proposed network was based on a multitasking heteroskedastic classification model (in which the variance of the outputs prior to activation is predicted), that is, each region of the tumor was treated as a binary classification problem. The patch size used was equal to 5$\times$196$\times$196 and the batch size equal to 2. They used a cross-validation of 5 iterations. The results obtained in each of the planes were averaged. The average Dice coefficients obtained were equal to 0.89, 0.83 and 0.81 for the whole tumor, the tumor core and the enhancing tumor core, respectively.

\subsubsection{Semantic segmentations}

A common aspect among many segmentation works was the realization of semantic segmentations. In the present work, the labeling was done voxel by voxel, that is, regardless of the size of the input patch, the output was a single label associated with the voxel located in the center of the patch. Since semantic segmentation labels a set of voxels at the same time, its implementation significantly reduces labeling time compared to voxel-by-voxel segmentation. However, in this work the reason for not having done a semantic segmentation was the following. Neural networks would have needed a larger amount of input data, as well as having a more complex architecture design, in order to develop the ability to identify the necessary patterns to correctly label a set of voxels at the same time. Due to the limitations of hardware and number of samples, a voxel-by-voxel segmentation was more feasible despite the increase in application time.

\subsubsection{Size of input patch and batch}

When large patch sizes are used, then the number of samples included in the batch is reduced. The reason is that there must be a balance between the size of the patches and the number of samples in the batch, due to space limitations in the memories of the computing systems (or the GPUs being precise). When a network uses large patches, more contextual information is included. But by reducing the batch size, then the variance increases during the gradient descent process. On the other hand, when a network uses small patches, contextual information is lost. However, increasing the number of samples in the batch improves optimization. In the present work, small patches (5$\times$5 or 15$\times$15 voxels) and large sample sizes in the batch (32 or 256) were used. Thus, training optimization and better use of available memory were preferred, although with the respective loss of contextual information.

\subsubsection{CNNs in 2D and 3D}

Regarding the contextual information of the patches, this is obviously greater when 3D patches are studied instead of 2D. However, there are some advantages to using 2D patches. For example, due to the different acquisition methods of MRI volumes in which the number of slices can change, a 3D segmentation model could not be equally applied for any volume. Thus a 2D model is more flexible for real applications \cite{mod_2d}. On the other hand, the algorithms are more efficient in time and memory when they analyze a smaller amount of information contained in the 2D patches. When 2D patches are studied they must be obtained from some plane of the MRI volumes. Unless there is justification for preferring one plane over the others, the usual way is to study all three axial, sagittal, and coronal planes. For all of the above, in this work three different CNNs were created by studying 2D patches for each of the three planes.

\subsubsection{Cascade segmentation}

There is a methodology called cascade segmentation. This means that the segmentation of the different regions of interest of the tumor are performed sequentially. This can be done starting from a rough segmentation of the whole tumor, and then make consecutive refinements until all the subregions of the tumor are segmented \cite{3_2018}. Another way is following a reverse hierarchical order, in which the region of least difficulty is segmented first (usually the whole tumor) and then the ones with the greatest difficulty \cite{cascade_1} are segmented consecutively. A different methodology is to segment the different regions of the tumor independently \cite{cascade_2}. This has the disadvantage that the training time can be longer and there is a risk of overfitting in each network because they are trained separately.
Since cascade segmentation is a sequential process, then errors in a previous segmentation affect subsequent segmentations. Despite this, in the present work, a cascade segmentation was performed with a reverse hierarchical order, starting with the whole tumor, then the core tumor, and finally the enhancing tumor core. At each stage, the best possible segmentation was attempted before continuing with the next. During stage 3 (the most difficult), the segmentations of stages 1 and 2 were studied separately, in addition to the entire volume. With this, it was possible to avoid that the errors of stage 2 (which preserved errors of stage 1) dominate over the segmentation of greater difficulty made in stage 3.

\subsubsection{Concatenations}

In the CNNS developed, concatenation operations between some layers were included. This was motivated by the so-called dense convolutional networks (DenseNet) \cite{densenet}. It has been shown that CNNs can be more accurate and efficient if they include shorter connections between the layers near the input and output. Based on this idea, each layer of a DenseNet is connected to all the layers of the network, so that feedback is generated between all of them. With this network, the problem known as gradient disappearance decreases, the propagation of the characteristics increases (implying greater and better training) and the number of parameters is reduced. On the other hand, the most used CNN called U-Net also includes in its architecture symmetric concatenations between the encoder and decoder sections to generate autocontext \cite{1,2,2_2018}. Unlike the DenseNet and the U-Net, in the CNNs of the present work only the outputs of two adjacent layers were connected to form the input of the next layer (Figs. \ref{red_1} and \ref{red_3_1}). A greater number of connections would have implied a greater computation time. However, with the proposed architecture, the advantage of feedback between layers was obtained.

\subsubsection{Post-processing}

Just as in the present work different post-processing techniques were applied, there are also other works that used some tools after applying a CNN. Some examples are the following. The so-called conditional random field (CRF) model in 3D, which reduces false positives and refines the segmentation \cite{crf, crf2, cnn_dens_del}. Application of a connected component analysis (CC) for the elimination of segmentation artifacts due to magnetic field inhomogeneities \cite{cc, cc2, cc3, cnn_dens_del}. A post-processing in which hierarchical information about regions of the tumor contained in each other is used, so that everything that was found outside the region of interest is eliminated \cite{out}. Processes to fill hallows and remove small clusters of voxels using thresholds obtained by comparing the ratio between the size of the region of interest and the possible region to remove \cite{holes_elim}. The use of DenseNet to improve delineation of segmented tumor, air, and brain \cite{cnn_dens_del}. Replacement of the enhancing tumor core by necrotic tissue when the volume of the first one was less than a certain threshold chosen individually for each glioma \cite{1_2019, 3_2019}.

Regarding the participants of the BraTS challenge of the years 2018 and 2019 \cite{lib_2018, lib_2019}, only a few explicitly reported having performed any post-processing, with CRF and CC being the most used techniques. A smaller amount of works dealt with the problem of eliminating the enhancing tumor core with a simple post-processing using thresholds, in addition to performing a filling process or elimination of small groups of erroneously labeled voxels. So, unlike other works where the segmentations depended mainly on the trained CNNs (or other types of models), in the present work different original post-processing techniques were the main responsible for correctly performing the segmentations and therefore obtain results at the level of the state of the art.

\subsubsection{Results comparison}

Some common characteristics among the works that obtained the best results in the BraTS challenge were the study of large 3D patches, architectures with numerous layers and filters, semantic segmentation and little or no post-processing of the segmentations made. In the present work, small 2D patches were studied, CNN architectures with relatively few layers were built, voxel-by-voxel labeling was performed, and various post-processing techniques were proposed, the latter being the main responsible for the results obtained. As shown in Table \ref{comp}, the average Dice coefficients calculated with the volumes tested in the segmentations made of the whole tumor, tumor core and the enhancing tumor core, satisfactorily reached the level of the state of the art defined by the winners of the BraTS challenge of the year 2019. It should be mentioned that a direct comparison cannot be strictly made between the results of the challenge and those obtained in the present work. The reason is that in both cases the same gliomas were not studied. However, although the gliomas in both cases were different, this is the best comparison that could be made.

\subsection{Limitations and future work}

Among the main limitations was the hardware available to carry out the work. As mentioned on several occasions, this prevented the proposal of a segmentation methodology with high computational demands such as those reported by other works. Assuming computing tools with better technical specifications were available, then future work would include creating CNNs for semantic segmentation, using larger 3D patches to extract contextual information, building more complex architectures, and applying transfer learning. Another limitation was the number of samples studied. The methodologies described should be applied to a larger number of samples, so that the generalization and reproducibility of the results reported here can be verified. For this, the necessary parameter values for each of the proposed processes must be refined, which could be done by studying other databases through independent work. Also, the study of a larger number of samples would be convenient to make a more adequate training of the RNN of stage 3. On the regression models $M_1$ $M_2$, their variables must be refined in such a way that only the ones with the greatest weight are preserved through the use of various mathematical methods designed for such a task. In stage 3, more complex processes were described than in stages 1 and 2 due to the difficulty in segmenting the enhancing tumor core. However, many of these processes were devised at the time of studying stage 3, so these were never applied in the first two stages. Something similar happened in stage 2 with respect to stage 1. Therefore, some processes applied in one stage could be implemented in another earlier stage. Since the results could depend on the validation volumes (although there was reproducibility with the training and testing volumes), other forms of validation should be considered such as cross-validation, ROC curves, etc. Also other MRI modalities could be studied for the segmentation of gliomas. In general, the applicability of the proposed post-processing techniques and processes in general could be tested in other segmentation problems.

\section{Conclusions}

An automatic glioma segmentation methodology was developed. Considering a cascade approach, a methodology was proposed to segment different histological subregions of gliomas in a hierarchical manner. The segmented subregions were the whole tumor, the tumor core, and the enhancing tumor core. Different CNNs were trained by analyzing conventional MRI to perform the segmentations. Due to hardware limitations, the CNNs had non-complex architectures and studied a small number of samples. Therefore, a set of original post-processing techniques was developed to improve the segmentations made by CNNs. With the testing volumes, average Dice coefficients equal to 0.8934, 0.8376, and 0.8113 were obtained for the subregions of the whole tumor, the tumor core, and the enhancing tumor core, respectively. These results reached the state of the art in the segmentation of gliomas, represented by the participants who ranked first in the BraTS challenge of the year 2019. Future work includes the application of the proposed methods to other databases with number of larger samples, to corroborate the results obtained and to refine the parameter values used in the post-processing techniques. Since the original methods and procedures presented in this work successfully obtained outstanding results, then the main goal is to be a useful contribution in the development of complementary tools in the field of clinical diagnosis and segmentation of gliomas.

\bibliographystyle{unsrt}
\bibliography{ref}

\end{document}